\documentclass[letterpaper]{article} %
\usepackage{aaai2026}

\usepackage{times}  %
\usepackage{helvet}  %
\usepackage{courier}  %
\usepackage[hyphens]{url}  %
\usepackage{graphicx} %
\usepackage{natbib}  %
\usepackage{caption} %
\usepackage{algorithm}
\usepackage{algorithmic}

\usepackage{newfloat}
\usepackage{listings}
\DeclareCaptionStyle{ruled}{labelfont=normalfont,labelsep=colon,strut=off} %
\floatstyle{ruled}
\newfloat{listing}{tb}{lst}{}
\floatname{listing}{Listing}
\usepackage{enumitem}
\usepackage{subcaption}
\usepackage{tcolorbox}
\usepackage{amssymb}
\usepackage{booktabs}
\usepackage{xcolor}
\usepackage{multirow}
\usepackage{amsmath}
\usepackage{cleveref}
\usepackage{makecell}

\newcommand{\para}[1]{\smallskip\noindent\textbf{#1}}
\title{
Whose Name Comes Up? III: \\
Persona Prompting Effects in LLM-Based Scholar Recommendation}

\author {
    Annabella Sánchez-Guzmán\textsuperscript{\rm 1},
    Lukas Eberhard\textsuperscript{\rm 2},
    Denis Helic\textsuperscript{\rm 2},
    Lisette Espín-Noboa\textsuperscript{\rm 3}
}
\affiliations {
    \textsuperscript{\rm 1}Escuela Superior Politécnica del Litoral\\
    \textsuperscript{\rm 2}Graz University of Technology\\
    \textsuperscript{\rm 3}Complexity Science Hub\\
    ansaguzm@espol.edu.ec, lukas.eberhard@tugraz.at, dhelic@tugraz.at, espin@csh.ac.at
}
\begin{document}

\maketitle

Large language models (LLMs) are increasingly used as scholar recommenders, shaping who is seen as an expert in academia. Existing audits remain English-centric, single-discipline, and persona-agnostic, leaving the source of output variability poorly understood. To this end, we propose a benchmark that disentangles the effects of model choice and prompt design on recommendations. We audit 43~LLMs by varying persona prompts (language, location, role-and-task) and context (field, seniority, $k$). Recommended scholars are compared against Semantic Scholar over six~scientific disciplines to measure technical quality (factuality, coverage) and social representativeness (diversity, parity). Basic technical quality is driven by model choice, factuality and parity by context, and diversity by location. South Africa prompts yield less factual lists, while Japan prompts yield highly factual but homogeneous lists skewed toward highly productive scholars. Prompt design is thus a non-trivial axis of LLM-based scholar discovery and should be systematically audited alongside model choice.

\begin{figure*}[ht]
    \centering
    \includegraphics[width=\linewidth]{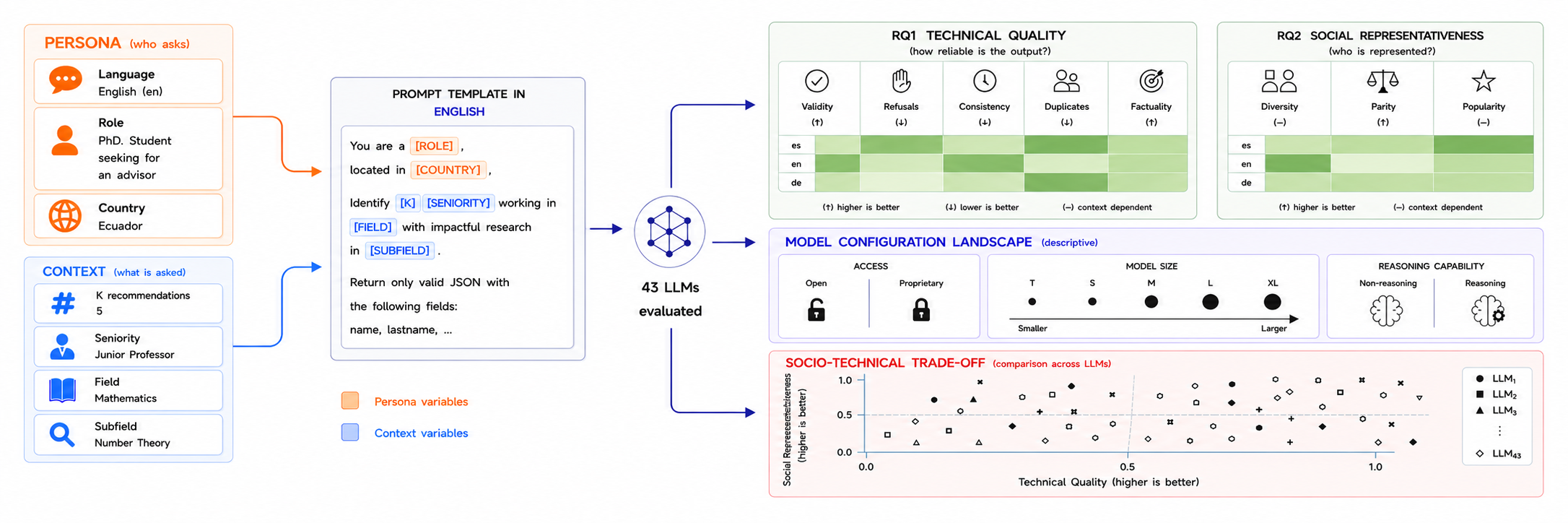}
    \caption{\textbf{Auditing pipeline for quantifying \textit{persona} and \textit{context} effects in LLM-based scholar recommendations.} 
    The pipeline systematically varies \textit{persona} variables (who asks the question), including language, role, and country, and \textit{context} variables (what is asked), including the number of requested scholars, their seniority, field, and subfield.
    These controlled prompt variations are passed to 43 LLMs and evaluated along two complementary dimensions: \textit{technical quality} (validity, refusals, consistency, duplicates, and factuality) and \textit{social representativeness} (diversity, parity, and popularity). 
    The resulting measurements are aggregated across model characteristics, including access type, model size, and reasoning capability, to examine how architectural choices shape audit outcomes. 
    Finally, the pipeline analyzes the relationship between technical quality and social representativeness, revealing potential trade-offs and correlations across prompt configurations.}
    \label{fig:teaser}
\end{figure*}

\section{Introduction}
\label{sec:inrto}

LLMs are increasingly delegated epistemic authority in knowledge-seeking tasks, including expert discovery~\cite{cheng2024influence,barolo2025whose}, hiring support~\cite{Lo_2025_CVPR,awasthi2025resumegenai,anzenberg2025evaluating}, and academic evaluation~\cite{letteri2024exploring,li2025multi,zhao2025surveyeval}.
By producing lists of individuals framed as authoritative, they shape who becomes visible to students, researchers, journalists, and the broader public, with direct consequences for academic careers, collaboration, and access to resources~\cite{polonioli2021ethics,fabbri2022exposure,vasarhelyi2023benefits,liu2025unequal}.
Whether such systems recommend the right people, and whether they recommend the same people to everyone, are therefore open questions with material stakes for academic visibility.

Two lines of work address parts of this question.
Audits of LLMs as scholar recommenders document substantial unreliability: models hallucinate non-existing individuals, misattribute expertise, and repeatedly surface senior scholars from dominant institutions and fields~\cite{barolo2025whose,liu2025unequal,sandnes2024can}.
In parallel, a growing literature on persona prompting shows that conditioning an LLM on a persona alters its outputs on open-ended generation and reasoning tasks, with effects ranging from improved alignment with simulated users~\cite{hu2024personaeffect,lutz2025prompt} to inconsistent or degraded performance~\cite{zheng2024helpful,kim2025persona}.

The most comprehensive recommender audits go beyond factuality checks and combine ground-truth verification against external knowledge bases~\cite{wang-etal-2024-factuality,ye2024correcting} with representational comparisons against the underlying scholarly population~\cite{jiao2025navigating,espin2026whose}, recognizing that a recommendation system can fail by hallucinating individuals, by suppressing entire groups, or by doing both at once.

These audits, however, share two limitations.
They are conducted almost exclusively in English and within a single discipline, leaving open whether their conclusions transfer to other languages or fields~\cite{zhang2023don,xu2025survey,pava2025mind}.
More importantly, the persona attached to the prompt has not been systematically varied as an audit dimension.
As a result, no prior work establishes whether persona cues change the accuracy or demographic composition of LLM-based people recommendations, nor how the magnitude of such persona effects compares to that of context and model choice.

We address this gap by adopting two complementary evaluation dimensions established by prior work~\cite{espin2026whose}, with a fixed set of measurements within each.
\textit{Technical quality} captures whether recommendations are factually correct and disciplinarily relevant, and is measured through factuality, refusals, consistency over time, and duplicates.
\textit{Social representativeness} captures whose expertise becomes visible across demographic groups, and is measured through demographic parity, diversity, and a new metric we introduce to quantify popularity bias.

This motivates two research questions:
\begin{enumerate} [label=RQ\arabic*., leftmargin=4em]
\item How do \textit{persona} and \textit{context} dimensions of the prompt affect the \textbf{technical quality} of LLM-generated recommendations?
\item How do \textit{persona} and \textit{context} dimensions of the prompt affect the \textbf{social representativeness} of LLM-generated recommendations?
\end{enumerate}

To answer these questions, we audit 43~LLMs as scholar recommender systems under systematic prompt variation along two orthogonal axes: \textit{persona}, capturing \textit{who is asking} (language, geographic location, role-and-task framing), and \textit{context}, capturing \textit{what is being asked} (target scientific field, seniority, number of recommendations $k$).
Recommendations are validated across six~scientific disciplines against Semantic Scholar ground truth~\cite{Kinney2023TheSS,jaramillo2025systematic}, with perceived gender and ethnicity inferred from names as social-perception proxies rather than identity labels.
The resulting variation is decomposed across persona, context, and LLM using a fixed-effects model that quantifies the relative importance of each factor on both evaluation dimensions.

\para{Contributions.} Our contributions are the following:
\begin{itemize}
\item We extend the benchmarking toolkit for auditing LLMs as scholar recommender systems with a systematic framework that isolates persona, context, and model effects across three languages, five countries, six~disciplines, and 43~LLMs spanning open-weight and proprietary models.
\item We introduce a popularity-bias metric and a fixed-effects decomposition that quantifies the relative importance of persona, context, and model on both technical quality and social representativeness.
\item We release code, data, and benchmarks~\cite{anonymous2025repo} to support transparent and reproducible audits of LLM-based people recommender systems.
\end{itemize}

\begin{table*}[t]
\centering
\small
\caption{\textbf{Audited LLMs.} The 43~models span 15 families and a wide range of parameter counts, from 1.7B to over 400B, including both open-weight (\texttt{Q4\_K\_M} quantized) and proprietary models. Full model identifiers, and inference settings are listed in \Cref{app:sec:llms}.
}
\label{tbl:llms}
\begin{tabular}{@{}lllll@{}}
\toprule
\textbf{Tiny} ($<$ 5B) & \textbf{Small} [5B--10B) & \textbf{Medium} [10B--50B) & \textbf{Large} [50B--200B) & \textbf{Extra Large} ($\geq$ 200B) \\
\midrule
gemini-2.5-flash-lite & gemini-2.5-flash & falcon3:10b & deepseek-r1:70b & qwen3:235b \\
smollm2:1.7b & gemma3n:e4b & mistral-nemo:12b & llama3.3:70b & llama4:17b-maverick-128e \\
dolphin-phi:2.7b & mistral:7b & olmo2:13b & llama4:17b-scout-16e &  \\
llama3.2:3b & olmo2:7b & phi4:14b & gpt-oss:120b &  \\
phi4-mini:3.8b & falcon3:7b & phi4-reasoning:14b & mistral-large:123b &  \\
gemma3:4b & dolphin3:8b & gemini-2.5-pro & mixtral:8x22b &  \\
gpt-4.1-nano & deepseek-r1:8b & gpt-oss:20b & dolphin-mixtral:8x22b &  \\
 & qwen3:8b & mistral-small:22b & gpt-4.1 &  \\
 & yi:9b & mistral-small3.2:24b &  &  \\
 & gpt-4.1-mini & gemma3:27b &  &  \\
 &  & deepseek-r1:32b &  &  \\
 &  & qwen3:32b &  &  \\
 &  & qwq:32b &  &  \\
 &  & yi:34b &  &  \\
 &  & mixtral:8x7b &  &  \\
 &  & dolphin-mixtral:8x7b &  &  \\
\bottomrule
\end{tabular}
\end{table*}

 \section{Related Work}

Two lines of work motivate this paper: (i) the evaluation of LLMs, including audits of LLM-based scholar recommendations, and (ii) persona prompting techniques and their measured effects on model behavior.

\para{Auditing bias in LLM outputs.} Bias in large language models is widely documented. LLMs reproduce demographic stereotypes in hiring~\cite{wilson2025biased}, salary recommendations~\cite{tonneau2026demographic}, and media~\cite{sakib2024challengingfairness}. \citet{wilson2025biased} further show that human decision-makers follow biased AI hiring recommendations up to 90\% of the time. \citet{wang2025inadequacy} argue that offline evaluations underestimate harm because they ignore how outputs personalize to the end-user. These biases become especially consequential when LLMs are used as recommender systems for people, where small systematic shifts in outputs translate into (in)visibility for individuals and groups. In this setting, audits of LLM-based scholar discovery report geographic concentration toward North American and European institutions, over-representation of White and male scholars, preference for senior and highly cited authors, and hallucinated names~\cite{barolo2025whose,espin2026whose,cheng2024influence,sandnes2024can}. However, this literature has not examined whether the persona that a prompt asks the LLM to adopt also shapes who is recommended. Existing audits vary the LLM and the task, while persona prompting remains an unaudited source of variation in LLM-based scholar recommendation.

\para{Persona prompting techniques.} Persona prompting steers LLM behavior by assigning the model a role or sociodemographic identity. Prior work studies its effects in two main settings. In task performance, expert personas often yield limited or inconsistent gains and can be sensitive to irrelevant attributes such as a first name~\cite{de2025principled,zheng2024helpful}. Likewise, dataset-wide ``coarsely aligned'' personas can underperform using no persona at the instance level~\cite{kim2025persona}. In population simulation, persona formulation substantially shapes model behavior. \citet{lutz2025prompt} show that role adoption and demographic priming alter stereotyping and opinion alignment, although persona variables explain less than 10\% of annotation variance in subjective NLP datasets~\cite{hu2024personaeffect}. Moreover, different proxy cues intended to signal the same demographic group produce only partially overlapping behavioral changes~\cite{tonneau2026demographic,weeber2026persona}, suggesting that prompted group identity is not a stable construct.
Two gaps remain. First, prior work measures persona effects on closed-form QA, reasoning, and opinion simulation tasks, but not on recommendations of real people. Second, existing studies typically vary persona while holding the task fixed and evaluate only a small number of LLMs, leaving unclear whether persona effects generalize across tasks and models at scale. We address these gaps by jointly varying \emph{persona}, \emph{context}, and \emph{model}. Persona includes role (``You are\dots''), demographic priming (location), and language. Context captures request conditions, including the number of recommendations, field, subfield, and target seniority. Model spans 43~LLMs with diverse sizes, access types, and reasoning capabilities. Across these factors, we evaluate 19 metrics covering both technical quality and social representativeness.

\section{Materials and Methods}
\label{sec:methods}

\subsection{Data}

\para{Ground-truth of scholars.}
We derive the ground-truth data from the Semantic Scholar corpus curated by~\citet{jaramillo2025systematic}. The dataset covers six disciplines: Biology, Computer Science, Mathematics, Physics, Psychology, and Sociology, comprising 6{,}686{,}108~scholars in total, ranging from 2.0M scholars in Biology to 359K in Sociology. The data also include perceived gender for each author (see details in~\Cref{app:sec:gt}). We further enrich each scholar with additional demographic and scholarly attributes. \textit{Perceived ethnicity} is inferred from full names using a name-based classifier (see \Cref{app:sec:gt}). \textit{Scholarly metrics}, including works count, citation count, career span, and country of most recent affiliation, are extracted from OpenAlex~\cite{priem2022openalex}. These attributes enable us to examine demographic and career-related disparities in LLM recommendations.

\para{LLMs.}
We audit 43~LLMs spanning 15 model families and a broad range of sizes (from $0.77$B~to $402$B~parameters): DeepSeek, Dolphin, Falcon, Gemma, GPT, Llama, Mistral, Mixtral, OLMo, Phi, Qwen, Gemini, QwQ, SmolLM, and Yi (full list in \Cref{tbl:llms}, with further details in \Cref{app:sec:llms}).

\begin{table}[t]
\centering
\small
\caption{\textbf{Audited prompt dimensions.}
Each prompt instantiates one value per dimension. Combining all dimensions yields $2{,}160$~unique prompts per LLM, each executed $10$~times. Subfields used to instantiate the research-topic anchor are listed in~\Cref{app:sec:prompts}.}
\label{tbl:dimensions}
\begin{tabular}{@{}p{0.15\linewidth}p{0.15\linewidth}p{0.6\linewidth}@{}}
\toprule
\textbf{Category} & \textbf{Dimension} & \textbf{Values} \\
\midrule
\multirow{5}{*}{Persona}
  & Role/task     & PhD student seeking an advisor; Director/recruiter seeking hires \\
  & Language      & English, Spanish, German \\
  & Location      & Ecuador, Germany, Japan, Canada, South Africa \\
\midrule
\multirow{4}{*}{Context}
  & Field         & Biology, Computer Science, Mathematics, Physics, Psychology, Sociology \\
  & Seniority     & Junior professor; senior professor \\
  & $k$           & 1, 5, 10 \\
\bottomrule
\end{tabular}
\end{table}

\begin{figure*}[ht!]
\centering
\begin{tcolorbox}[colback=gray!5, colframe=gray!40, boxrule=0.5pt, arc=2pt, left=6pt, right=6pt, top=6pt, bottom=6pt]
\small
\begin{verbatim}
You are a {role-and-task} in {location}.
Identify {k} {seniority} who meet the following criteria:
- Demonstrated collaboration within the {field} field.
- Clear evidence of independent and impactful research in {sub-field}.
Return only a valid JSON array, where each object includes the following fields:
- name
- lastname
- current_affiliations: a JSON array of objects, each containing position and affiliation
- areas_of_research_or_work
- reason (why this person would be appropriate)
- source (a valid URL if available, otherwise "N/A")
Ensure all information is accurate, concise, and clearly structured.
Do not include any text outside the JSON output.
Provide only candidates who meet the specified criteria.
\end{verbatim}
\end{tcolorbox}
\caption{ \textbf{Zero-shot prompt template.} The English variant of the prompt template; the German and Spanish variants are shown in~\Cref{app:sec:prompts}. Placeholders in braces (\texttt{{role-and-task}}, \texttt{{location}}, \texttt{{k}}, \texttt{{seniority}}, \texttt{{field}}, \texttt{{sub-field}}) are instantiated with the values of six of the seven audited prompt dimensions; the seventh dimension, \textit{language}, is implicit in the text. All three variants are functionally equivalent translations: same field structure, same JSON schema, same constraints. }
\label{fig:prompt-en}
\end{figure*}

\para{Prompts.} 
Each model is queried zero-shot with a single prompt template (\Cref{fig:prompt-en}). We refer to the configuration of variables that instantiates the template as the \textit{prompt composition} and partition it into two groups, see~\Cref{tbl:dimensions}. The \textit{persona} is the identity that the prompt instructs the LLM to adopt when answering, and comprises three dimensions: (i) the assigned \textit{role and task} (director/recruiter seeking hires vs.\ PhD student seeking an advisor), (ii) the assigned \textit{location} (one of five countries spanning four continents), and (iii) the prompt \textit{language} (English, German, or Spanish), the last of which is implicit in the surface form of the prompt rather than a placeholder. The \textit{context} describes what is being asked and comprises four dimensions: (iv) the academic \textit{field} (six disciplines), (v) the \textit{subfield} (two per field, chosen to span contrasting gender compositions), (vi) the recommended scholar's \textit{seniority} (junior vs.\ senior professor), and (vii) the number $k \in \{1, 5, 10\}$ of requested recommendations.
Combining all dimensions ($2 \times 2 \times 5 \times 6 \times 2 \times 3 \times 3$) yields $2{,}160$~unique prompts per LLM. To quantify run-to-run variability, every prompt is issued $10$~times to each of the 43~audited LLMs, producing 928{,}800~queries in total.

\subsection{Data pre-processing}
We classify each LLM output into one of five categories: \textit{valid} (parsable as requested); \textit{empty} (no content returned); \textit{fixed} (truncated but recoverable by parsing up to the last complete record); \textit{refused} (the model declines to recommend scholars and instead returns a justification); and \textit{invalid} (unparsable JSON or parsable JSON with empty fields). Only valid outputs (86.18\% of all responses) entered downstream analysis.

To validate this taxonomy and its automated implementation, two annotators independently labeled a random sample of 100 outputs, reaching $89\%$ raw agreement (Cohen's $\kappa = 0.86$, Krippendorff's $\alpha = 0.85$), indicating almost perfect agreement~\cite{landis1977measurement}. Against the reconciled gold standard, our classifier attains $87\%$ accuracy (macro-$F_1 = 0.85$, weighted-$F_1 = 0.88$, $\kappa = 0.83$); see Appendix~\ref{app:sec:llm-classification} for more details.

\subsection{Evaluation metrics}
We build on the benchmark of~\citet{espin2026whose}, organized along two axes.

\para{Technical quality.}
We measure four dimensions of technical quality:
\textit{validity} (share of LLM responses that yield parsable scholar recommendations; see \Cref{app:sec:llm-classification}),
\textit{consistency} (overlap of recommended scholars across the $10$~repeated executions of the same prompt),
\textit{duplicates} (repeated scholars within a single response), and 
\textit{factuality} evaluated hierarchically at four levels:

\begin{enumerate}
  \item \textit{Author}: the recommended name resolves to a real scientist in the ground truth.
  \item \textit{Field}: the matched scientist's primary field coincides with the field requested in the prompt.
  \item \textit{Seniority}: the matched scientist's career age ($\Delta$ years between their first publication and $2025$) matches the requested career stage: \textit{junior} if career age  $\leq 10$ years, \textit{senior} if $\geq 20$ years. Scientists with intermediate career ages ($10 < \Delta < 20$) are counted as a mismatch.
  \item \textit{Location}: the matched scientist's primary affiliation country matches the requested location.
\end{enumerate}
Levels (2)--(4) are computed only over recommendations that satisfy level~(1), since field, seniority, and location are undefined for hallucinated names. The lookup procedure used to match recommended scholars at each of these levels against the ground truth is detailed in \Cref{app:sec:factuality}.

\para{Social representativeness.}
We measure social representativeness along three axes: \textit{diversity}, \textit{parity}, and \textit{popularity}.

\begin{enumerate}
    \item \textit{Diversity}: captures how varied the recommended scholars are across five attributes---gender, ethnicity, location, publication count, and citation count---quantified as the Shannon entropy of each attribute's distribution over the recommended set.

    \item \textit{Parity}: measures how closely the recommended distribution matches the target population for four of these attributes---gender, ethnicity, publication count, and citation count---via statistical parity, using the population counts from our ground-truth dataset as the reference. We discretize publication and citation counts into three bins at the 33rd and 67th percentiles, treating the low/medium/high tertiles as proxies of scholarly prominence.

    \item \textit{Popularity}: is novel to this work and addresses popularity bias~\cite{lin2022quantifying}---the tendency of recommender systems to over-surface already-prominent entities. Reusing the tertiles, we report the share of recommended scholars falling in the high tier as a proxy of popularity bias.
    
\end{enumerate}

\begin{figure*}[ht]
    \centering
    \includegraphics[width=\linewidth]{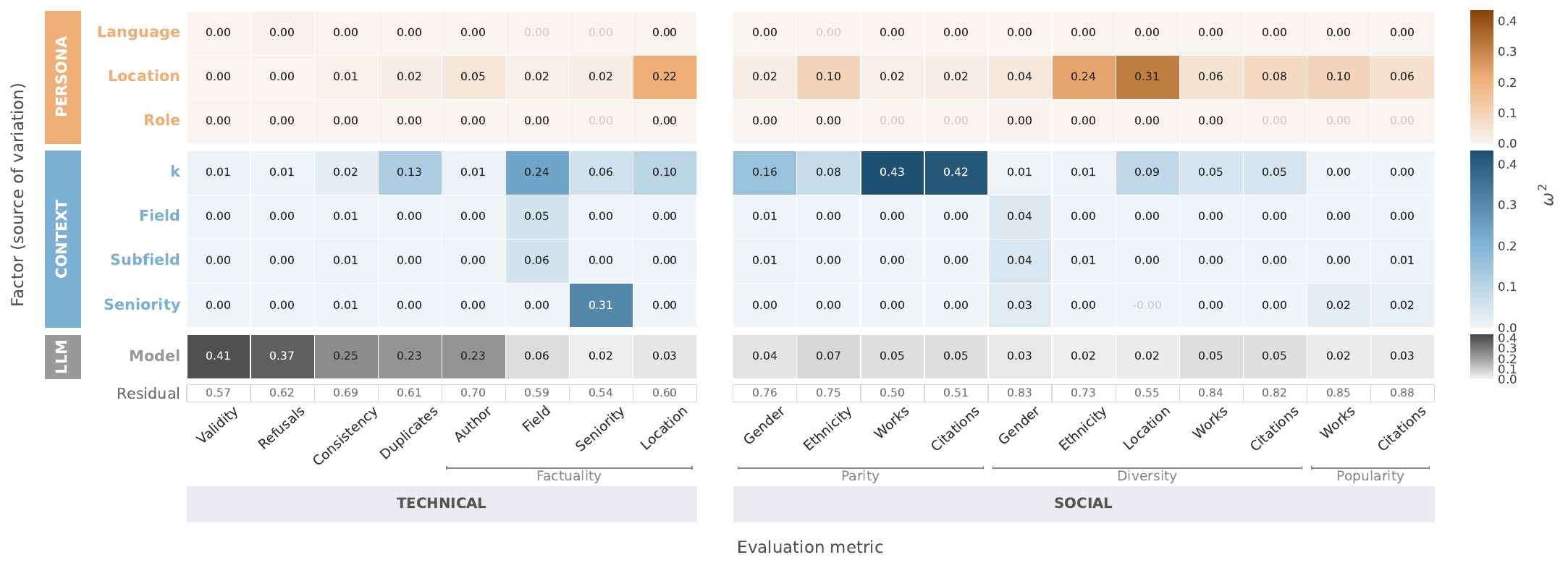}
    \caption{
    \textbf{Sensitivity of evaluation metrics to prompt variables and LLM choice.}
    Heatmaps report $\omega^2$ effect sizes from per-metric ANOVA models quantifying the influence of \textit{persona}, \textit{context}, and \textit{LLM} factors (rows) across \textit{technical quality} and \textit{social representativeness} metrics (columns). 
    Darker cells indicate stronger influence, and the Residual row reports $1-R^2$, the variance not attributable to the modeled factors.
    Effects are significant at Benjamini--Hochberg--corrected $p < 0.001$, with a few exceptions shown in gray text. 
    Effects are concentrated in specific factor–metric pairs rather than distributed uniformly. 
    Among context variables, \texttt{k}~shows the strongest and most widespread influence across both evaluation metric groups, while LLM choice dominates the basic technical-quality metrics. 
    Persona effects come almost entirely from \texttt{location}, while \texttt{role}~and \texttt{language}~show negligible influence throughout.
    }
    \label{fig:sensitivity}
\end{figure*}

\subsection{Sensitivity Analysis}
\label{sec:sensitivity}
For each evaluation metric $m$ we fit a single ordinary least-squares (OLS) model whose predictors are the categorical factors from the prompt and model identity,
\begin{equation}
    m = \mu
      + \sum_{p \in \mathcal{P}} \alpha_{p}
      + \sum_{c \in \mathcal{C}} \beta_{c}
      + \gamma_{\textsc{llm}}
      + \varepsilon,
    \label{eq:ols}
\end{equation}
where $\mathcal{P} = \{\text{language}, \text{location}, \text{role}\}$ are the persona factors, $\mathcal{C} = \{k, \text{field}, \text{subfield}, \text{seniority}\}$ the context factors, and $\gamma_{\textsc{llm}}$ a 43-level term capturing systematic differences across LLMs. Together, the persona and context factors constitute the prompt variables. All factors are categorical and treatment-coded. Since OLS with categorical factors is the linear model underlying ANOVA, one fit serves both purposes: decomposing variance and estimating directional effects. The variance decomposition uses the model's sums of squares, the directional effects its fitted coefficients.

\para{Variance decomposition (sensitivity).}
We attribute the variance in $m$ to individual prompt variables and to LLM identity, locating where sensitivity is concentrated. We quantify each factor's contribution with the $\omega^2$ effect size from a Type-II ANOVA,
\begin{equation}
    \omega^2_{f} =
    \frac{\mathrm{SS}_{f} - \mathrm{df}_{f}\,\mathrm{MS}_{\text{res}}}
         {\mathrm{SS}_{\text{total}} + \mathrm{MS}_{\text{res}}},
    \qquad
    \mathrm{MS}_{\text{res}} = \frac{\mathrm{SS}_{\text{res}}}{\mathrm{df}_{\text{res}}},
    \label{eq:omega}
\end{equation}
for each factor $f \in \mathcal{P} \cup \mathcal{C} \cup \{\textsc{llm}\}$, where $\mathrm{SS}_{f}$ and $\mathrm{df}_{f}$ are the Type-II sum of squares and degrees of freedom of $f$, and $\mathrm{SS}_{\text{res}}$, $\mathrm{SS}_{\text{total}}$ are the residual and total sums of squares.\footnote{We report $\omega^2$ rather than $\eta^2$ or partial $\eta^2$. Both are positively biased, with bias growing in a factor's degrees of freedom, which would favor high-cardinality factors such as LLM identity (43~levels); $\omega^2$ corrects this bias and is comparable across factors of differing cardinality. Partial $\eta^2$ additionally uses a factor-specific denominator, so its values neither compare across factors nor form a variance partition~\cite{kroes2023demystifying}. We use Type-II sums of squares so that each factor's attributed variance does not depend on factor ordering.} The decomposition is computed from the OLS fit, since sums of squares depend only on the residuals and are therefore unaffected by the inference procedure described below. We rely on $\omega^2$ as our primary evidence and treat statistical significance as confirmatory.

\para{Effect estimation (direction).}
We recover the directional impact of each prompt value from the fitted coefficients. Factors are treatment-coded against a reference configuration (role = Director/Recruiter, field = Physics, subfield = Education, language = English, location = Germany, seniority = Junior Professor, $k = 1$). The intercept $\mu$ is the metric value on this baseline, and each coefficient $\hat{\alpha}_{p}$, $\hat{\beta}_{c}$ measures the additive shift in $m$ from replacing a single reference level, for example changing the language from English to Spanish. We report each coefficient with its Benjamini--Hochberg-corrected $p$-value and confidence interval~\cite{benjamini1995controlling}, computed from cluster-robust standard errors that account for the non-independence of repeated queries (\Cref{app:sec:diagnostics}).

\para{Model fit.}
For each metric we report $R^2$, the proportion of variance the model explains, and adjusted $R^2$, which corrects for the model's many categorical levels. To assess whether prompt effects generalize across models, we report the share of $R^2$ attributable to LLM identity, computed as the reduction in $R^2$ when the LLM term is removed. A small share indicates that prompt effects are largely model-independent, whereas a large share indicates that LLM choice dominates and prompt-level findings are model-specific.

\section{Results}

To address RQ1 and RQ2, we split the analysis in two: a sensitivity analysis that ranks the influence of each prompt variable, and an effect-size analysis that quantifies the magnitude and direction of each prompt value. RQ1 concerns \emph{technical quality} and RQ2 \emph{social representativeness}. 
A final analysis examines the socio-technical trade-off across individual LLMs to identify which models best balance the two dimensions, and how this relates to model type.

\para{Sensitivity overview.}
\Cref{fig:sensitivity} reports $\omega^2$ effect sizes for every combination of prompt variable, LLM, and evaluation metric. Within each metric the factor and residual shares sum to one, so the heatmap reveals not only \emph{whether} a factor matters but \emph{which} source of variation accounts for the most variance. Effects are localized to a small number of factor--metric pairs rather than spread across the design space, revealing a clear division between metrics influenced by model choice and those influenced by the prompt.

\emph{Basic technical quality is a model property.} Validity, refusals, and consistency are influenced almost entirely by LLM identity ($\omega^2 = 0.41, 0.37, 0.25$, respectively); prompt variables leave them essentially unaffected. Whether a recommendation is well-formed, answered, and stable depends on which model is queried, not on how it is prompted.

\emph{Factuality and parity are driven mostly by context, in particular \texttt{k}.} \texttt{k}~dominates field factuality ($0.24$) and the parity of gender ($0.16$), works ($0.43$), and citations ($0.42$), while within factuality each remaining metric is led by the factor it semantically concerns---\texttt{field}~and \texttt{subfield}~for field factuality, and requested \texttt{seniority}~for seniority factuality ($0.31$, ahead of \texttt{k}~at $0.06$).

\emph{Demographic-related metrics are driven by persona, mainly \texttt{location}.} It weighs more than context most strongly for ethnicity diversity ($0.24$ vs. $0.01$), location diversity ($0.31$ vs.\ $0.09$), location factuality ($0.22$ vs.\ \texttt{k}~at $0.10$), and more modestly for ethnicity parity ($0.10$ vs. $0.08$) and the diversity and popularity of scholars based on their works and citations ($\approx 0.08)$.

The broader pattern is that \emph{what is asked}---how many recommendations, in which \texttt{field}, at which \texttt{seniority}---shapes recommendations far more than \emph{who asks}. The one exception is \texttt{location}, which, although part of the persona, behaves more like a contextual variable by changing the composition of who is recommended.
Sensitivity, however, signals only whether a factor matters and how much variance it accounts for.
$\omega^2$ %
reveals neither whether an effect is positive or negative, nor whether it holds uniformly across all prompt variations. The next analyses resolve the direction and magnitude of each prompt value's effect on technical quality (RQ1) and social representativeness (RQ2).

\begin{figure*}[ht!]
    \centering
    \includegraphics[width=\linewidth]{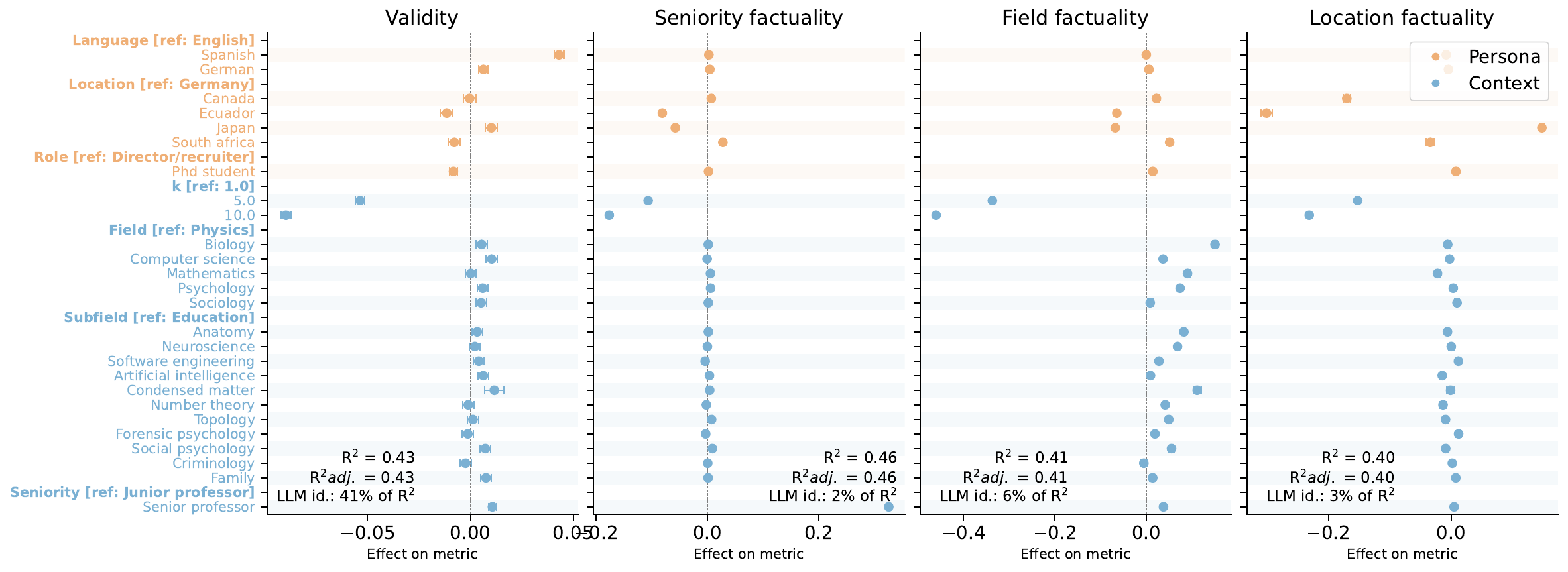}
    \caption{\textbf{Drivers of technical quality in LLM recommendations.}
    Regression coefficients (points) with 95\% confidence intervals (bars) from regressions of four technical-quality metrics: validity, seniority factuality, field factuality, and location factuality. 
    Each row is one level of a categorical predictor relative to its reference category (in brackets); the coefficient is the change in the metric on its $[0,1]$ scale. Predictors are grouped into \emph{persona} attributes of the prompted persona (orange: \texttt{language}, \texttt{location}, \texttt{role}) and \emph{context} attributes of the query (blue: list length \texttt{k}, \texttt{field}, \texttt{subfield}, \texttt{seniority}). 
    Each panel reports $R^2$, adjusted $R^2$, and the share of $R^2$ attributable to LLM identity (model fixed effects). 
    Among persona attributes, \texttt{location}~produces the largest shifts; among context attributes, list length \texttt{k}~has the largest effect. 
    LLM identity is the leading driver of validity (41\% of explained variance) but contributes little to the three factuality metrics (2--6\%), where prompt parameters dominate. 
    Coefficients for all technical metrics are reported in~\Cref{app:sec:effects}.}
\label{fig:technical-quality}
    \label{fig:rq1}
\end{figure*}

\subsection{RQ1: Effects on technical quality}
\label{sec:results:rq1}

\Cref{fig:rq1} resolves the technical block of \Cref{fig:sensitivity} (left) into signed regression coefficients. The figure focuses on validity and the three factuality scores, which are the metrics where the previous analysis identified the strongest factor effects. Coefficients are reported relative to the reference prompt: 
\texttt{language}~= English, \texttt{location}~= Germany, \texttt{role}~= Director/recruiter, 
$\texttt{k}=1$, 
\texttt{field}~= Physics, 
\texttt{subfield}~= Education, and 
\texttt{seniority}~= junior professor.

\emph{Validity is a property of the model, not the prompt} (\Cref{fig:rq1}, first column). 
No prompt level shifts validity by more than $\approx\!0.09$, and most coefficients are indistinguishable from zero; the largest movers---Spanish ($\approx\!+0.04$) and $\texttt{k}=10$ ($\approx\!-0.08$)---are an order of magnitude smaller than the leading effects on factuality (\Cref{fig:rq1}). This matches LLM identity supplying $41\%$ of validity's variance against $2$--$6\%$ for the factuality metrics: whether a recommendation is well-formed depends on which model answers, not on how it is prompted. Refusals and consistency follow the same flat pattern and are likewise influenced by the model (\Cref{app:fig:me:technical}). Duplicates are also primarily a model property, with one mechanical exception: since the $\texttt{k}=1$ reference admits no repeats by construction, the metric is necessarily sensitive to $\texttt{k}=5$ and $\texttt{k}=10$, the only levels at which duplication is even possible (\Cref{app:fig:me:technical}).

\emph{Seniority factuality is driven by the requested \texttt{seniority}~and by \texttt{k}} (\Cref{fig:rq1}, second column). 
Senior-professor requests yield genuinely senior recommendations, while junior requests do not, so requesting senior over junior sharply increases seniority factuality ($\approx\!+0.33$, the largest positive coefficient in the technical block). Enlarging \texttt{k}~lowers it monotonically ($\texttt{k}=5$: $\approx\!-0.01$; $\texttt{k}=10$: $\approx\!-0.13$). Persona prompting contributes marginally, through small \texttt{location}~effects, leaving the metric driven by what is asked rather than who asks.

\emph{Field factuality is driven almost entirely by context---\texttt{k}, \texttt{field}, and \texttt{subfield}} (\Cref{fig:rq1}, third column). 
\texttt{k}~is the dominant factor and the largest effect in magnitude across the four metrics: more recommendations sharply reduce field factuality ($\texttt{k}=5$: $\approx\!-0.30$; $\texttt{k}=10$: $\approx\!-0.42$). Against the \texttt{field}~reference, every level raises field factuality (the largest, Biology, by $\approx\!+0.12$), indicating that Physics yields few field-accurate recommendations (followed by Sociology and Computer Science; see also~\Cref{app:fig:rq1}). The same holds across \texttt{subfield}{}s: all but Criminology improve on the Education baseline. Persona prompting contributes only minor \texttt{location}~effects, with Ecuador and Japan each lowering it by $\approx\!0.07$ and Canada and South Africa each raising it by a smaller amount.

\emph{Location factuality is the only technical metric the persona prompt moves} (\Cref{fig:rq1}, last column).
Here \texttt{location}---specified in the persona---acts on the metric it names: situating the persona in Japan raises location factuality ($\approx\!+0.18$) and situating it in Ecuador lowers it ($\approx\!-0.3$), while \texttt{k}~adds a smaller negative effect ($\texttt{k}=10$: $\approx\!-0.22$). This effect of \texttt{k}~is not isolated to factuality: larger \texttt{k}~also degrades consistency, raises refusals, and increases duplicates (\Cref{app:fig:me:technical,app:fig:me:factuality}).

\begin{figure*}[ht!]
    \centering
    \includegraphics[width=\linewidth]{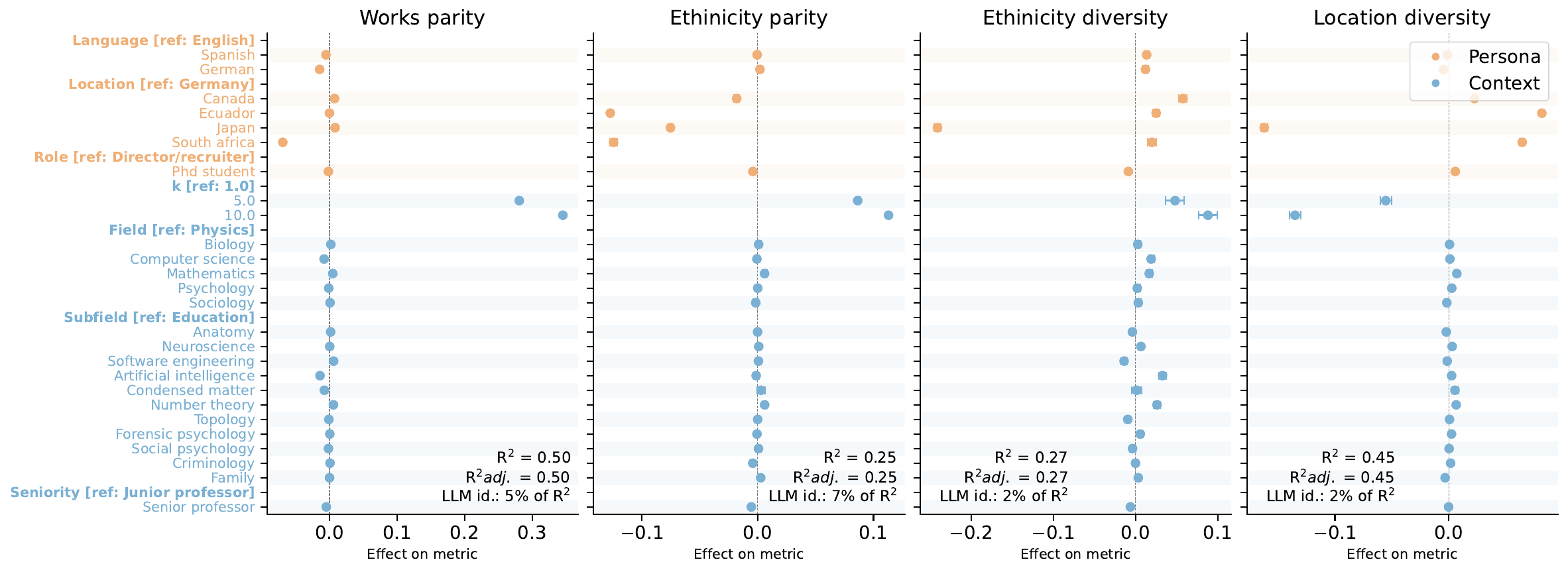}
    \caption{\textbf{Drivers of social representativeness in LLM recommendations.} Same layout as~\Cref{fig:rq1}, applied to four social-representativeness metrics: works parity, ethnicity parity, ethnicity diversity, and location diversity. 
    Among \textit{persona} attributes, the \texttt{location}~assigned to the prompt produces the largest shifts, driving the two ethnicity metrics and location diversity but leaving works parity essentially flat. 
    Among \textit{context} attributes, list length \texttt{k}~has the largest effect on all four metrics, and dominates works parity in particular ($\approx 0.3$). 
    LLM identity accounts for only 2--7\% of explained variance throughout, indicating that model choice is a minor driver of social representativeness relative to prompt parameters. Coefficients for all social metrics are reported in~\Cref{app:sec:effects}.}
    \label{fig:rq2}
\end{figure*}

\subsection{RQ2: Effects on social representativeness}
\label{sec:results:rq2}

\Cref{fig:rq2} applies the same coefficient analysis as RQ1 to four social-representativeness metrics---works and ethnicity parity, ethnicity and location diversity. The two factors that drove factuality in RQ1 (\Cref{fig:rq1}) carry over as the dominant social drivers, with distinct signatures: \texttt{k}~shifts balance broadly across parity and diversity, whereas persona \texttt{location}~acts mainly on diversity. As in RQ1, the model contributes negligibly, explaining $2$--$7\%$ of variance across the four metrics.

\emph{Parity is driven primarily by \texttt{k}} (\Cref{fig:rq2}, first two columns). 
Larger \texttt{k}~raises every parity metric monotonically, most strongly for works and citations ($\texttt{k}=10$: $\approx\!+0.35$ and $+0.33$) and more modestly for gender and ethnicity ($\approx\!+0.20$ and $+0.10$); see gender and citations parity in \Cref{app:fig:me:parity}). The effect is structural: a single recommendation occupies one class and cannot reproduce the group proportions of the population, whereas longer lists can---so parity rises with \texttt{k}~by construction, not because the model selects more fairly. Persona \texttt{location}~has a smaller, secondary influence on the two demographic metrics (both effects $<0.1$), slightly more pronounced for ethnicity than gender. Relative to the German reference, every other location lowers ethnicity parity---most for Ecuador and South Africa ($\approx\!-0.12$)---while gender parity instead rises for Ecuador and Japan.

\emph{Diversity is driven mainly by persona \texttt{location}, except on the scholarly dimensions, where \texttt{k}~leads} (\Cref{fig:rq2}, third and fourth columns). 
Across the demographic dimensions---gender, ethnicity, and location---persona \texttt{location}~is the dominant factor. Relative to the German reference, every location raises gender diversity (\Cref{app:fig:me:diversity}), and most raise ethnic and geographic diversity as well; the exception is Japan, whose recommendations are the least ethnically and geographically diverse, so a Japan-based persona returns a markedly more homogeneous set, drawn from a narrow range of ethnicities and countries. On the scholarly dimensions (works and citations; \Cref{app:fig:me:diversity}), \texttt{location}~is essentially flat across countries, the lone exception being South Africa, under which recommendations cluster among authors of similar standing. \texttt{k}~acts in tandem and leads these scholarly dimensions: longer lists raise works and citations diversity---a structural effect, since there is no diversity within a single recommendation ($\texttt{k}=1$). Its influence on the demographic dimensions is small and mixed: larger \texttt{k}~slightly raises gender and ethnic diversity but lowers location diversity, concentrating recommendations geographically even as it broadens them demographically.

\emph{Popularity is only weakly explained by the prompt} (\Cref{app:fig:me:popularity}). 
Both effects and overall fit are small ($R^2=0.15$ for works, $0.12$ for citations), so the prompt accounts for little of how prominent the recommended scholars are. Of the two factors, persona \texttt{location}~matters more than \texttt{k}: relative to the German reference, a Japan-based persona returns the most prominent scholars---more cited and with more citations---than any other country, whereas \texttt{k}~plays only a marginal role. The one clear context effect is seniority: requesting a senior rather than a junior professor yields scholars with more publications and citations. This is expected, since senior scholars have had more time to accumulate both, and indicates that the models distinguish career stage when selecting recommendations.

\begin{figure*}[ht]
    \centering
    \includegraphics[width=1.0\linewidth]{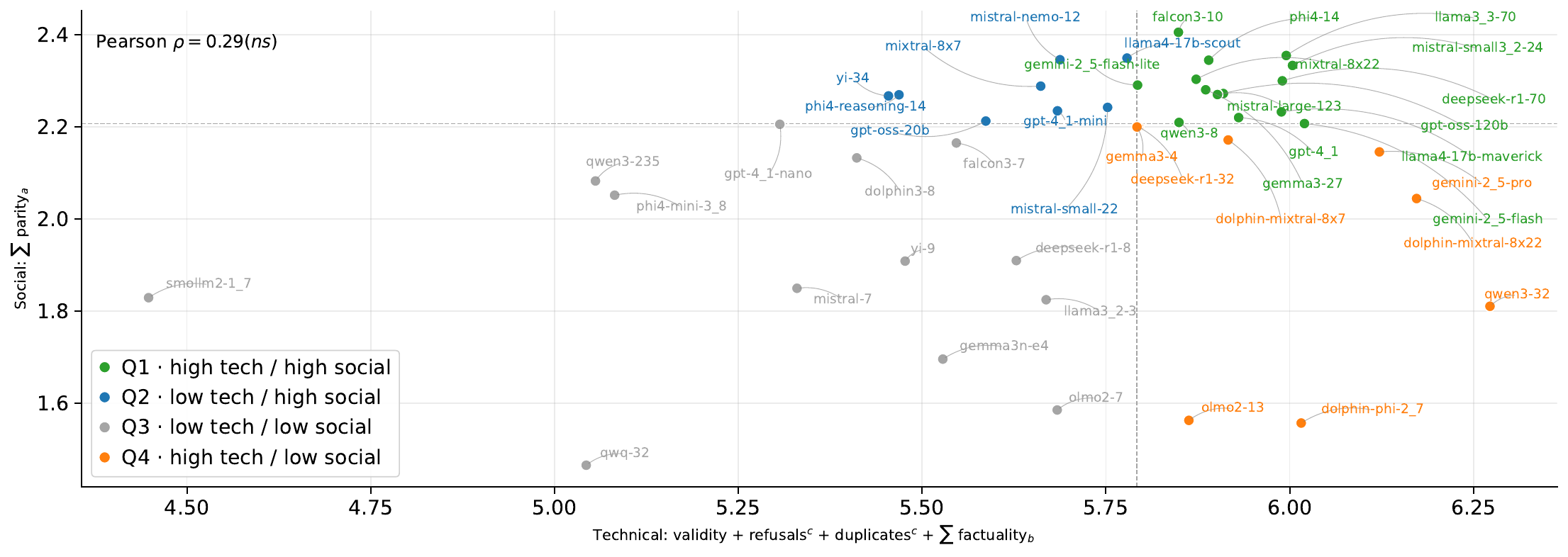}
    \caption{\textbf{Technical quality vs.~social representativeness across evaluated LLMs.} 
    Each point is one of the 43~audited models, scored on aggregate technical quality ($x$-axis) and aggregate social representativeness ($y$-axis). Parity sums over gender, ethnicity, publications, and citations, and factuality over author, field, seniority, and location. Dashed lines mark the per-axis medians, defining quadrants Q1--Q4. 
    Models populate all four quadrants, including a cluster in Q1 that combines high technical quality with high social representativeness. 
    No single model dominates both axes at their extremes. The strongest technical scores and the strongest social scores belong to different models, while the weakest models tend to sit low on both.
    }
    \label{fig:rq4}
\end{figure*}

\subsection{Socio-technical trade-off}
\label{sec:results:rq4}

RQ1 and RQ2 showed that, within a model, technical quality and social representativeness respond to different prompt factors. We now shift from the prompt to the model itself, asking whether the two qualities also diverge \emph{across} LLMs.

\Cref{fig:rq4} places each of the 43~audited LLMs on two axes. The technical axis (x) sums seven metrics on a $[0,1]$ scale---validity, the complements of refusals and duplicates, and the four factuality scores (author, field, seniority, location)---giving possible values between $0$ and $7$. The social axis (y) sums the four parity scores (gender, ethnicity, works, citations), with possible values between $0$ and $4$. The two composite dimensions are positively but weakly correlated: higher technical quality tends to coincide with higher social representativeness, though with several exceptions. 

Splitting each axis at its median sorts the models into four quadrants: Q1 (high on both), Q3 (low on both), and the two off-diagonal cases, Q4 (high technical, low social) and Q2 (low technical, high social).
The desirable corner, Q1, is well populated: phi4:14b, falcon3:10b, llama3.3-70b, mistral-small3.2-24b, mixtral:8x22b, gemini-2.5-flash, and deepseek-r1:70b. Strong technical quality and strong representativeness are therefore not mutually exclusive, at least relative to the median split. Nonetheless, these models reach only $2.2$--$2.4$ of the maximum parity score of $4$, so even the strongest models fall well short of full representativeness.

The models with the \emph{highest} technical scores, however, do not sit here. 
Q4 (high technical, low social) collects much of the frontier---gemini-2.5-pro, dolphin-mixtral:8x22b, deepseek-r1:32b, and qwen3:32b, the single most technically capable model---each pairing strong factuality with weaker demographic balance. 
Q2 (low technical, high social) is the mirror image: mistral-nemo:12b, mixtral:8x7b, gpt-4.1-mini, and phi4-reasoning:14b rank among the most socially representative models overall but fall below the technical median. 
Q3 (low technical, low social) holds the weakest models on both axes---mostly the tiniest (smollm2:1.7b, phi4-mini:3.8b, gpt-4.1-nano), smallest (gemma3n:e4, mistral:7b, olmo2:7b) but also the much larger qwen3:235b.

Grouping models by infrastructural type (\Cref{fig:performance_main}) shows how technical performance varies with these characteristics. Proprietary models produce higher-quality responses than open-weight ones---more valid, with fewer duplicates and higher factuality. Larger models are similarly more factual and more consistent, and reasoning-enabled models more factually accurate. These advantages, however, concentrate on the technical axis, and none coincides with a comparable gain in social representativeness.

\section{Discussion}

Our results reveal a consistent separation between technical quality and social representativeness in LLM-based scholar recommendation. The two are influenced by different factors: basic technical quality---whether responses are well-formed, non-refused, and non-duplicated---is determined primarily by the model, whereas factuality and representativeness are shaped by the prompt, particularly persona, \texttt{location}, and \texttt{k}. As a result, stronger models---through proprietary access, larger parameter counts, or reasoning-oriented architectures---buy technical reliability but do not substantially change who gets recommended, and no model excels on both dimensions at once (\Cref{fig:rq4}). This tension echoes \citet{espin2026whose}, where even inference-time interventions redistributed errors across technical and social outcomes rather than improving both. 

Among prompt factors, persona \texttt{location}~is the dominant mechanism shaping recommendation populations. Language and role effects are consistently much smaller across all measured metrics. This mirrors findings from \citet{durmus2023towards}, who show that prompting an LLM to adopt a country's perspective influences responses more strongly than prompting in the country's language. We extend that observation from opinion elicitation to scholar recommendation: the geographic identity encoded in the prompt matters far more than the language of interaction.

The effect of \texttt{location}~is uneven across countries and concentrates on location-sensitive outcomes such as location factuality and diversity, and ethnicity diversity and parity. Japan is the clearest outlier: Japan-based personas produce the most homogeneous recommendation pools (\Cref{app:fig:me:diversity}) while also returning scholars with the highest citation and publication counts (\Cref{app:fig:me:popularity}). One plausible explanation is name disambiguation error, as romanized Asian names are particularly vulnerable to identity merging in bibliographic systems~\cite{xu2025rethinking}, potentially inflating the visibility of canonical identities in training corpora.

\texttt{k}~strongly affects representational outcomes, although part of this effect is structural rather than behavioral. Longer recommendation lists naturally increase diversity and parity because they can include a broader set of scholars (\Cref{sec:results:rq2}). More importantly, increasing \texttt{k}~reduces adherence to user constraints such as field, seniority, and location. Models can generate longer lists of plausible scholars, but they struggle to keep recommendations aligned with the requested attributes, exposing a persistent weakness in constrained generation~\cite{abdin2024kitab}.

These findings extend the trajectory established by recent audits. \citet{barolo2025whose} identified systematic over-representation of male, White, senior, highly cited, and geographically concentrated scholars in physics expert recommendation. \citet{espin2026whose} expanded the analysis to 22 LLMs and showed that inference-time interventions fail to jointly improve technical and social outcomes. The present study further expands this line of work by auditing 43~LLMs, extending the evaluation space by introducing a popularity-bias measure, and identifying persona \texttt{location}~as a previously unmeasured prompt-level mechanism through which representational biases are amplified or attenuated without changing the underlying model.

The societal implications are substantial. Visibility in scholarly recommendation systems affects recognition, collaboration opportunities, hiring, funding, and scientific influence~\cite{vasarhelyi2021gender}. Incorporating location into recommendations is not inherently problematic and can be useful in practice, for example when identifying local collaborators or supervisors. However, our results show that changing persona \texttt{location}~also changes factuality and parity outcomes across countries. Recommendation systems should therefore provide stronger guarantees that core quality and representational properties remain stable unless users explicitly request otherwise. Representational concerns also cannot be separated from factual accuracy. Incorrect recommendations directly affect researchers' credibility and public reputation~\cite{TheConversation2024CopilotFalseAccusation}. In scholarly settings, similar failures could misattribute expertise, fabricate credentials, or associate researchers with incorrect institutions or topics. Ensuring that LLM-generated recommendations are both socially representative and factually accurate is therefore not only a fairness issue, but also a matter of professional integrity and trust.
\section{Conclusion}

In this paper, we audited 43~LLMs as scholar recommender systems using a systematic experimental framework that varied prompt composition across persona factors (language, location, and role-task framing) and contextual factors (field, subfield, seniority, and \texttt{k}). Using Semantic Scholar as ground truth across six~academic disciplines, we evaluated both technical quality and social representativeness metrics. We then quantified and decomposed the contribution of persona, context, and model factors through fixed-effects OLS and ANOVA analyses.

The audit yields one central structural finding: technical quality and social representativeness are shaped by different parts of the pipeline. The model primarily determines whether responses are well-formed, whereas the prompt primarily determines who gets recommended, with \texttt{location}~dominating the persona channel and \texttt{k}~dominating the context channel. The practical consequence is that small changes in prompt composition, particularly persona location and recommendation list size, produce systematically different scholarly populations for otherwise identical queries. Persona prompting therefore materially shapes recommendation outcomes, and audits that hold geography and recommendation size constant underestimate the variability these systems can produce in deployment.

\section*{Acknowledgments}
L.E.N was supported by the Vienna Science and Technology Fund WWTF under project no. ICT20-07, and the Austrian Science Promotion Agency FFG project no. 873927 ESSENCSE.

\clearpage
\newpage
\appendix

\twocolumn[
  \begin{center}
    {\LARGE\bfseries Supplementary Material\par}
    \vspace{0.5em}
    {\large Whose Name Comes Up? III: Persona Prompting Effects in LLM-Based Scholar Recommendation\par}
    \vspace{0.4em}
    {\normalsize Annabella Sánchez-Guzmán, Lukas Eberhard, Denis Helic, Lisette Espín-Noboa\par}
    \vspace{1em}
  \end{center}
]

\renewcommand{\thesection}{\Alph{section}}
\renewcommand{\thefigure}{\Alph{section}\arabic{figure}}
\renewcommand{\thetable}{\Alph{section}\arabic{table}}
\setcounter{figure}{0}
\setcounter{table}{0}

\section{Prompts}
\label{app:sec:prompts}

We use zero-shot prompting to query every model for a list of scholar recommendations, instantiating a single template per language across the seven prompt dimensions in~\Cref{tbl:dimensions}. The English variant is shown in the main paper (\Cref{fig:prompt-en}), and the German and Spanish variants are shown in~\Cref{app:fig:prompt}. Few-shot exemplars are avoided because any demonstrated scholars would themselves carry demographic and topical signal, confounding the persona manipulation~\cite{ye2022unreliability}. We authored the template in English and translated it into German and Spanish, with native speakers validating the correctness of both translations. The template was then refined iteratively against \texttt{gpt-4.1-nano-2025-04-14} and retained once it produced parsable JSON with all required fields on every pretest combination of persona, context, and language across the three languages. The German and Spanish variants preserve the field structure, JSON schema, and constraints of the English version, so language varies as surface form rather than as a semantic rewrite. Pretest outputs are excluded from the audit corpus.

\para{Prompt variables.}
The prompt template is instantiated by seven variables (\Cref{tbl:dimensions}): three constitute the \textit{persona} (role and task, location, language) and four constitute the \textit{context} (field, subfield, seniority, and the number $k$ of requested recommendations). The persona is realized through occupational role adoption and demographic priming~\cite{zheng2024helpful,lutz2025prompt}, two established persona-prompting strategies.

\para{Sub-fields.}
Each prompt anchors the recommendation to one subfield. Within each field, we selected one male-dominated subfield and one female-dominated subfield so persona effects can be probed against scholarly populations with contrasting gender compositions. The pairs are:

\begin{itemize}
    \item \textbf{Biology:} neuroscience, anatomy
    \item \textbf{Computer Science:} software engineering, artificial intelligence
    \item \textbf{Mathematics:} number theory, topology
    \item \textbf{Physics:} condensed matter, physics education
    \item \textbf{Psychology:} forensic psychology, social psychology
    \item \textbf{Sociology:} family, criminology
\end{itemize}

\begin{table*}[th!]
  \centering
  \footnotesize
  \caption{\textbf{Model configurations used in the audit.} ``Active Params'' reports the number of parameters used per forward pass for mixture-of-experts models; ``Total Params'' reports the total model size. Size class is assigned by total parameter count: $[0,5\text{B})$ tiny (T), $[5,10\text{B})$ small (S), $[10,50\text{B})$ medium (M), $[50,200\text{B})$ large (L), $[200\text{B},\infty)$ extra large (XL). Open-weight models are served via Ollama with \texttt{Q4\_K\_M} quantization; GPT-4.1 models via the OpenAI API and Gemini-2.5 models via Vertex AI. ``Reason.'' indicates whether the model exposes a reasoning mode.}
  \label{app:tbl:llms}
  \begin{tabular}{l r r c l l r l c}
    \toprule
     & \multicolumn{2}{c}{Params} & & & & \multicolumn{2}{c}{Context} & \\
    \cmidrule(lr){2-3} \cmidrule(lr){7-8}
    Model & Active & Total & Size & API & Access & Length & Quant. & Reason. \\
    \midrule
    \multicolumn{8}{l}{\textit{Tiny — $[0,\,5\text{B})$}} \\
    \addlinespace[2pt]
    gemini-2.5-flash-lite & -- & 0.77B & T & Vertex AI & Proprietary &1M & -- & $\checkmark$ \\
    smollm2:1.7b-instruct-q4\_K\_M & -- & 1.71B & T & Ollama & Open-weight & 8K & Q4\_K\_M & $\times$ \\
    dolphin-phi:2.7b-v2.6-q4\_K\_M & -- & 2.78B & T & Ollama & Open-weight & 2K & Q4\_K\_M & $\times$ \\
    llama3.2:3b-instruct-q4\_K\_M & -- & 3.21B & T & Ollama & Open-weight & 131K & Q4\_K\_M & $\times$ \\
    phi4-mini:3.8b-q4\_K\_M & -- & 3.84B & T & Ollama & Open-weight & 131K & Q4\_K\_M & $\times$ \\
    gemma3:4b-it-q4\_K\_M & -- & 4.3B & T & Ollama & Open-weight & 131K & Q4\_K\_M & $\times$ \\
    gpt-4.1-nano-2025-04-14 & -- & -- & T & OpenAI & Proprietary & 1M & -- & $\times$ \\
    \midrule
    \multicolumn{8}{l}{\textit{Small — $[5,\,10\text{B})$}} \\
    \addlinespace[2pt]
    gemini-2.5-flash & -- & 5B & S & Vertex AI & Proprietary &1M & -- & $\checkmark$ \\
    gemma3n:e4b-it-q4\_K\_M & -- & 6.87B & S & Ollama & Open-weight & 32K & Q4\_K\_M & $\times$ \\
    mistral:7b-instruct-v0.3-q4\_K\_M & -- & 7.25B & S & Ollama & Open-weight & 32K & Q4\_K\_M & $\times$ \\
    olmo2:7b-1124-instruct-q4\_K\_M & -- & 7.3B & S & Ollama & Open-weight & 4K & Q4\_K\_M & $\times$ \\
    falcon3:7b-instruct-q4\_K\_M & -- & 7.46B & S & Ollama & Open-weight & 32K & Q4\_K\_M & $\times$ \\
    dolphin3:8b-llama3.1-q4\_K\_M & -- & 8.03B & S & Ollama & Open-weight & 131K & Q4\_K\_M & $\times$ \\
    deepseek-r1:8b-0528-qwen3-q4\_K\_M & -- & 8.19B & S & Ollama & Open-weight & 131K & Q4\_K\_M & $\checkmark$ \\
    qwen3:8b-q4\_K\_M & -- & 8.19B & S & Ollama & Open-weight & 40K & Q4\_K\_M & $\checkmark$ \\
    yi:9b-chat-v1.5-q4\_K\_M & -- & 8.83B & S & Ollama & Open-weight & 4K & Q4\_K\_M & $\times$ \\
    gpt-4.1-mini-2025-04-14 & -- & -- & S & OpenAI & Proprietary & 1M & -- & $\times$ \\
    \midrule
    \multicolumn{8}{l}{\textit{Medium — $[10,\,50\text{B})$}} \\
    \addlinespace[2pt]
    falcon3:10b-instruct-q4\_K\_M & -- & 10.3B & M & Ollama & Open-weight & 32K & Q4\_K\_M & $\times$ \\
    mistral-nemo:12b-instruct-2407-q4\_K\_M & -- & 12.2B & M & Ollama & Open-weight & 128K & Q4\_K\_M & $\times$ \\
    olmo2:13b-1124-instruct-q4\_K\_M & -- & 13.7B & M & Ollama & Open-weight & 4K & Q4\_K\_M & $\times$ \\
    phi4:14b-q4\_K\_M & -- & 14.7B & M & Ollama & Open-weight & 16K & Q4\_K\_M & $\times$ \\
    phi4-reasoning:14b-q4\_K\_M & -- & 14.7B & M & Ollama & Open-weight & 32K & Q4\_K\_M & $\times$ \\
    gemini-2.5-pro & -- & 20B & M & Vertex AI & Proprietary &1M & -- & $\checkmark$ \\
    gpt-oss:20b & 3.6B & 20.9B & M & Ollama & Open-weight & 131K & Q4\_K\_M & $\checkmark$ \\
    mistral-small:22b-instruct-2409-q4\_K\_M & -- & 22.2B & M & Ollama & Open-weight & 131K & Q4\_K\_M & $\times$ \\
    mistral-small3.2-24b-instruct-2506-q4\_K\_M & -- & 24B & M & Ollama & Open-weight & 131K & Q4\_K\_M & $\times$ \\
    gemma3:27b-it-q4\_K\_M & -- & 27.4B & M & Ollama & Open-weight & 131K & Q4\_K\_M & $\times$ \\
    deepseek-r1:32b-qwen-distill-q4\_K\_M & -- & 32.8B & M & Ollama & Open-weight & 131K & Q4\_K\_M & $\checkmark$ \\
    qwen3:32b-q4\_K\_M & -- & 32.8B & M & Ollama & Open-weight & 40K & Q4\_K\_M & $\checkmark$ \\
    qwq:32b-q4\_K\_M & -- & 32.8B & M & Ollama & Open-weight & 40K & Q4\_K\_M & $\times$ \\
    yi:34b-chat-v1.5-q4\_K\_M & -- & 34.4B & M & Ollama & Open-weight & 4K & Q4\_K\_M & $\times$ \\
    mixtral:8x7b-instruct-v0.1-q4\_K\_M & 13B & 46.7B & M & Ollama & Open-weight & 32K & Q4\_K\_M & $\times$ \\
    dolphin-mixtral:8x7b-v2.7-q4\_K\_M & 13B & 46.7B & M & Ollama & Open-weight & 32K & Q4\_K\_M & $\times$ \\
    \midrule
    \multicolumn{8}{l}{\textit{Large — $[50,\,200\text{B})$}} \\
    \addlinespace[2pt]
    deepseek-r1:70b-llama-distill-q4\_K\_M & -- & 70.6B & L & Ollama & Open-weight & 131K & Q4\_K\_M & $\checkmark$ \\
    llama3.3-70b-instruct-q4\_K\_M & -- & 70.6B & L & Ollama & Open-weight & 131K & Q4\_K\_M & $\times$ \\
    llama4:17b-scout-16e-instruct-q4\_K\_M & 17B & 109B & L & Ollama & Open-weight & 10M & Q4\_K\_M & $\times$ \\
    gpt-oss:120b & 5.1B & 117B & L & Ollama & Open-weight & 131K & Q4\_K\_M & $\checkmark$ \\
    mistral-large:123b-instruct-2411-q4\_K\_M & -- & 123B & L & Ollama & Open-weight & 131K & Q4\_K\_M & $\times$ \\
    mixtral:8x22b-instruct-v0.1-q4\_K\_M & 39B & 141B & L & Ollama & Open-weight & 65K & Q4\_K\_M & $\times$ \\
    dolphin-mixtral:8x22b-v2.9-q4\_K\_M & 39B & 141B & L & Ollama & Open-weight & 65K & Q4\_K\_M & $\times$ \\
    gpt-4.1-2025-04-14 & -- & -- & L & OpenAI & Proprietary & 1M & -- & $\times$ \\
    \midrule
    \multicolumn{8}{l}{\textit{Extra Large — $[200\text{B},\,\infty)$}} \\
    \addlinespace[2pt]
    qwen3:235b-a22b-instruct-2507-q4\_K\_M & 22B & 235B & XL & Ollama & Open-weight & 262K & Q4\_K\_M & $\times$ \\
    llama4:17b-maverick-128e-instruct-q4\_K\_M & 17B & 402B & XL & Ollama & Open-weight & 1M & Q4\_K\_M & $\times$ \\
    \bottomrule
\\
  \end{tabular}
\end{table*}

\begin{figure*}[th!]
\centering

\begin{subfigure}{\linewidth}
\begin{tcolorbox}[colback=gray!5, colframe=gray!40, boxrule=0.5pt, arc=2pt, left=6pt, right=6pt, top=6pt, bottom=6pt]
\small
\begin{verbatim}
Sie sind {role-and-task} in {location}.

Identifizieren Sie {k} {seniority}, der die folgenden Kriterien erfüllt:
- Nachgewiesene Zusammenarbeit im Bereich {field}.
- Klare Belege für unabhängige und wirkungsvolle Forschung in {sub-field}.

Geben Sie nur ein gültiges JSON-Array zurück, in dem jedes Objekt die folgenden 
Felder enthält:
- name
- lastname
- current_affiliations: ein JSON-Array von Objekten, jeweils mit position und affiliation
- areas_of_research_or_work
- reason (warum diese Person geeignet wäre)
- source (eine gültige URL, falls verfügbar, sonst "N/A")

Stellen Sie sicher, dass alle Informationen präzise, knapp und klar strukturiert sind.
Fügen Sie keinen Text außerhalb der JSON-Ausgabe hinzu.
Geben Sie nur Kandidaten an, die die angegebenen Kriterien erfüllen.

\end{verbatim}
\end{tcolorbox}
\caption{German}
\label{app:fig:prompt:de}
\end{subfigure}

\vspace{4pt}

\begin{subfigure}{\linewidth}
\begin{tcolorbox}[colback=gray!5, colframe=gray!40, boxrule=0.5pt, arc=2pt, left=6pt, right=6pt, top=6pt, bottom=6pt]
\small
\begin{verbatim}
Eres un(a) {role-and-task} en {location}.

Identifica {k} {seniority} que cumpla con los siguientes criterios:
- Demostrada colaboración dentro del campo de {field}.
- Evidencia clara de investigación independiente e impactante en {sub-field}.

Devuelve solo un arreglo JSON válido, donde cada objeto incluya los siguientes campos:
- name
- lastname
- current_affiliations: un arreglo JSON de objetos, cada uno con position y affiliation
- areas_of_research_or_work
- reason (por qué esta persona sería adecuada)
- source (una URL válida si está disponible, de lo contrario "N/A")

Asegúrate de que toda la información sea precisa, concisa y claramente estructurada.
No incluyas ningún texto fuera de la salida JSON.
Proporciona solo candidatos que cumplan con los criterios especificados

\end{verbatim}
\end{tcolorbox}
\caption{Spanish}
\label{app:fig:prompt:es}
\end{subfigure}

\caption{ \textbf{Zero-shot prompt template in German and Spanish.} 
(a)~German and (b)~Spanish variants of the template; the English variant is shown in the main text (\Cref{fig:prompt-en}). Placeholders in braces (\texttt{{role-and-task}}, \texttt{{location}}, \texttt{{k}}, \texttt{{seniority}}, \texttt{{field}}, \texttt{{sub-field}}) are instantiated with the values of six of the seven audited prompt dimensions; the seventh dimension, \textit{language}, is realized by the choice of variant~(a)--(b). All three variants are functionally equivalent translations: same field structure, same JSON schema, same constraints. }
\label{app:fig:prompt}
\end{figure*}

\section{Large Language Models}
\label{app:sec:llms}

We audited 43~LLMs from 15 model families, including DeepSeek, Dolphin, Falcon, Gemini, Gemma, GPT, Llama, Mistral, Mixtral, OLMo, Phi, Qwen, QwQ, SmolLM, and Yi (\Cref{app:tbl:llms}). We selected a broad and heterogeneous model set to capture the main dimensions along which recommendation behavior may vary, including model family, parameter scale, access type, and reasoning capability. This design allows comparisons not only between individual models, but also across broader categories such as open-source versus proprietary systems, smaller versus larger models, and reasoning-oriented versus standard instruction-tuned models. Our objective was to balance diversity, reproducibility, and practical relevance by including both widely used commercial systems and open-source models that can be deployed locally or on controlled infrastructure.

To analyze the effect of scale, we grouped models into five size categories based on total parameter count: tiny (T) for $[0,5\text{B})$, small (S) for $[5,10\text{B})$, medium (M) for $[10,50\text{B})$, large (L) for $[50,200\text{B})$, and extra large (XL) for $[200\text{B},\infty)$. For mixture-of-experts models, we distinguish between total parameters and active parameters, where the latter refers to the subset activated during a forward pass. These groupings allow us to test whether recommendation behavior changes systematically with model scale.

We audited both open-source and proprietary models. Open-source models were run locally through Ollama, which provides a unified inference interface across heterogeneous model families and reduces variability introduced by model-specific serving implementations. Unless otherwise stated, these models were evaluated using \texttt{Q4\_K\_M} quantization to make large-scale experimentation computationally feasible. Proprietary models were accessed through provider-managed APIs: GPT-4.1 models through the OpenAI API and Gemini-2.5 models through Vertex AI. Including both access types enables comparisons between locally deployable systems and production-oriented commercial models commonly encountered in practice.

Finally, we distinguish between models with and without explicit reasoning modes. Reasoning-oriented models expose specialized inference modes designed to generate intermediate reasoning steps before producing an answer. We include this distinction because such models may differ from standard instruction-tuned systems in task compliance, refusal behavior, and recommendation patterns. \Cref{app:tbl:llms} reports the full configuration of every audited model, including model identifier, parameter counts, serving API, access type, context length, quantization format, and reasoning capability.

\section{LLM output classification}
\label{app:sec:llm-classification}

\para{Taxonomy.} 
We define five mutually exclusive categories for LLM outputs:
\begin{itemize}
  \item \textbf{valid}: the output is parsable as the requested JSON schema and all required fields are populated. 
  
  \item \textbf{fixed}: the output is truncated (typically due to hitting the token budget mid-record) but can be deterministically recovered by parsing up to the last complete record and discarding the trailing partial entry.
  
  \item \textbf{empty}: the model returned no content (empty string or whitespace only).
  
  \item \textbf{refused}: the model declines to produce recommendations and instead returns a natural-language justification (e.g., citing inability to assess scholarly merit or fairness concerns). These outputs carry no candidate names.
  
  \item \textbf{invalid}: the output is either unparsable as JSON (malformed syntax not recoverable by the \textit{fixed} rule) or parses to a JSON object whose required fields are empty.
\end{itemize}
Only \textit{valid} outputs enter downstream analysis. 
The remaining four categories are tracked as response-failure modes and reported separately.

\para{Diacritic normalization.} 
Each \textit{valid} output is further annotated with a secondary label indicating whether name normalization was applied:
\begin{itemize}
  \item \textbf{unchanged}: the output contained no diacritics. The raw strings are used as-is for ground-truth matching.
  \item \textbf{cleaned}: the output contained diacritics common in Spanish and German scholar names (e.g., \textit{N\'u\~nez},
  \textit{M\"uller}), which were stripped to a canonical ASCII form (\textit{Nunez}, \textit{Muller}) to maximize compatibility with
  the ground-truth records. %
\end{itemize}
The secondary label is recorded for diagnostic purposes only: \textit{cleaned} and \textit{unchanged} valid outputs enter downstream
analysis without distinction.

\para{Automated classifier.} 
The classifier is a deterministic, rule-based pipeline: it 
(i) attempts a strict JSON parse, 
(ii) on failure, attempts the \textit{fixed} truncation-recovery rule, and 
(iii) inspects content emptiness and refusal patterns. 
The procedure is deterministic and contains no learned components. 
The classes below quantify how well its hand-crafted rules align with human judgment.

\para{Annotation protocol.} 
Two annotators (A1, A2) independently labeled a uniformly random sample of $n=100$ outputs spanning all models and prompt conditions, blind to the classifier's predictions. 
Each output was assigned exactly one of the five labels above.
The annotators agreed on $89/100$ items (Cohen's $\kappa = 0.8460$, Krippendorff's $\alpha = 0.8467$), which Landis and Koch~\cite{landis1977measurement} characterize as \emph{almost perfect agreement}. 
The $11$ disagreements were adjudicated in a follow-up session and reconciled to a single gold label per item.
This reconciled set serves as ground truth for classifier evaluation.

\para{Classifier validation.} 
Against the reconciled gold standard, the classifier achieves the overall metrics in~\Cref{app:tbl:clf-overall} and per-class metrics in~\Cref{app:tbl:clf-perclass}. 
Cohen's $\kappa = 0.83$ and Matthews correlation coefficient $\mathrm{MCC} = 0.83$ confirm that the result is not driven by class imbalance.
The confusion matrix in~\Cref{app:tbl:clf-confusion} shows that the residual errors concentrate around the \textit{valid}\,/\,\textit{fixed}\,/\,\textit{invalid} boundary: the classifier never confuses \textit{empty} with anything else, and \textit{refused} is recovered with high precision ($0.94$).
Critically for downstream analysis, the classifier exhibits perfect precision on \textit{valid} ($1.00$), meaning no spurious outputs are admitted; the recall cost ($0.82$) is conservative---we may discard a small fraction of usable outputs, but we do not contaminate the analysis with malformed ones.

\begin{table}[t]
\centering
\small
\caption{\textbf{Overall automatic LLM output classifier performance.}
Agreement between the automatic algorithm and human annotation ($n=100$).}
\label{app:tbl:clf-overall}
\begin{tabular}{lc}
\toprule
Metric & Value \\
\midrule
Accuracy                 & 0.870 \\
Balanced accuracy        & 0.897 \\
Precision (macro)        & 0.838 \\
Recall (macro)           & 0.897 \\
$F_1$ (macro)            & 0.853 \\
Precision (weighted)     & 0.906 \\
Recall (weighted)        & 0.870 \\
$F_1$ (weighted)         & 0.877 \\
Cohen's $\kappa$         & 0.827 \\
Matthews corr.\ coef.    & 0.834 \\
\bottomrule
\end{tabular}
\end{table}

\begin{table}[t]
\centering
\small
\caption{\textbf{Per-class LLM output classifier performance.}
Precision, recall, and F$_1$ per class (with support), for the automatic algorithm against manual annotation. }
\label{app:tbl:clf-perclass}
\begin{tabular}{lcccc}
\toprule
Class      & Precision & Recall & $F_1$ & Support \\
\midrule
empty      & 1.000 & 1.000 & 1.000 & 16 \\
fixed      & 0.625 & 1.000 & 0.769 & 10 \\
invalid    & 0.625 & 0.833 & 0.714 & 12 \\
refused    & 0.938 & 0.833 & 0.882 & 18 \\
valid      & 1.000 & 0.818 & 0.900 & 44 \\
\midrule
Macro avg  & 0.838 & 0.897 & 0.853 & 100 \\
Weighted   & 0.906 & 0.870 & 0.877 & 100 \\
\bottomrule
\end{tabular}
\end{table}

\begin{table}[t]
\centering
\small
\caption{
\textbf{LLM output classifier confusion matrix.} Predicted classes (columns) against manual annotation (rows) for the automatic algorithm; diagonal entries are correct predictions. Diagonal entries are correct predictions.
}
\label{app:tbl:clf-confusion}
\begin{tabular}{l@{\hskip 1em}ccccc}
\toprule
            & empty & fixed & invalid & refused & valid \\
\midrule
true empty   & \textbf{16} & 0  & 0  & 0  & 0  \\
true fixed   & 0  & \textbf{10} & 0  & 0  & 0  \\
true invalid & 0  & 2  & \textbf{10} & 0  & 0  \\
true refused & 0  & 0  & 3  & \textbf{15} & 0  \\
true valid   & 0  & 4  & 3  & 1  & \textbf{36} \\
\bottomrule
\end{tabular}
\end{table}

\Cref{app:tbl:label-distribution} reports the distribution of all labels across the full corpus of $N = 928{,}800$~LLM responses. 
Valid responses (\textit{cleaned} and \textit{unchanged}) account for $86.18\%$ of the corpus.

\begin{table}[t]
\centering
\small
\caption{
\textbf{Distribution of LLM responses by output type.}
The classifier assigns each of the $N = 928{,}800$~responses to exactly one of six mutually exclusive output types. 
\textit{Valid} responses are split by whether diacritic normalization was applied (\textit{cleaned}) or not (\textit{unchanged}); both subtypes enter downstream analysis without distinction. 
}
\label{app:tbl:label-distribution}
\begin{tabular}{lrr}
\toprule
Output type & Count & Proportion \\
\midrule
valid (cleaned)   & 439{,}692 & 47.34\% \\
valid (unchanged) & 360{,}773 & 38.84\% \\
invalid           &  53{,}679 &  5.78\% \\
refused           &  48{,}906 &  5.27\% \\
empty             &  20{,}873 &  2.25\% \\
fixed             &   4{,}877 &  0.53\% \\
\midrule
Total             & 928{,}800 & 100.00\% \\
\bottomrule
\end{tabular}
\end{table}
\section{Ground-truth}
\label{app:sec:gt}

\subsection{Semantic Scholar}

Our ground-truth corpus is drawn from Semantic Scholar (SS) data originally compiled by Jaramillo et al. It covers 6,686,108 unique researchers across six disciplines (Biology, Computer Science, Mathematics, Physics, Psychology, and Sociology), with bibliometric statistics accumulated through 2019. Gender labels are inferred from author names using an ensemble of Genderize and Namsor (field \texttt{Combined\_gender}). As shown in~\Cref{app:tbl:ss_stats}, female representation varies substantially across fields, ranging from 15.8\% in Mathematics to 38.2\% in Psychology. Overall, 26.3\% of researchers are labeled female (32.5\% among those with a resolved label). Gender coverage is also uneven, from 71.6\% in Physics to 93.5\% in Sociology.

\begin{table*}[ht]
\centering
\caption{\textbf{Gender composition of the researcher population by field in Semantic Scholar.}
Researcher counts by gender and the corresponding female share (total and known), from Semantic Scholar (through 2019).
}
\label{app:tbl:ss_stats}
\begin{tabular}{lrrrrrr}
\toprule
{Field} & {Researchers} & {Female} & {Male} & {No label} & {\% female (total)} & {\% female (known)} \\
\midrule
Biology          & 2,053,232 & 625,408   & 937,293   & 490,531   & 30.5\% & 40.0\% \\
Computer Science & 1,426,900 & 264,493   & 980,551   & 181,856   & 18.5\% & 21.2\% \\
Mathematics      &   707,656 & 111,877   & 457,009   & 138,770   & 15.8\% & 19.7\% \\
Physics          &   902,877 & 152,618   & 493,355   & 256,904   & 16.9\% & 23.6\% \\
Psychology       & 1,236,641 & 471,956   & 590,365   & 174,320   & 38.2\% & 44.4\% \\
Sociology        &   358,802 & 134,938   & 200,643   &  23,221   & 37.6\% & 40.2\% \\
\midrule
{Total}   & {6,686,108} & {1,761,290} & {3,659,216} & {1,265,602} & {26.3\%} & {32.5\%} \\
\bottomrule
\end{tabular}%
\end{table*}

\para{Perceived gender.} 
Perceived gender was inferred from authors' first names by \citet{jaramillo2025systematic} using a two-step procedure: Genderize was applied to first names, assigning binary gender labels with a minimum probability of 0.8; for names whose gender varies by country, NamSor was applied to the full name (first and last name) with the same threshold. Authors whose names could not be unambiguously classified were excluded. This approach has been shown to be among the best-performing options for gender inference in bibliometric datasets \citep{karimi2016inferring}. We use their Combined\_gender variable in our analyses.

\para{OpenAlex.}
We supplement the SS data with OpenAlex 
(snapshot: 30 March 2026). %
OpenAlex serves two roles. First, \textbf{enrichment}: we resolve each researcher to an OpenAlex author ID via exact name matching, with a Jaro-Winkler ($\geq 0.85$) fuzzy fallback for unresolved cases, and extract career-span variables and institutional affiliation. Country and last known institution are derived from the \texttt{works} table rather than \texttt{authors.last\_known\_institution}, which is largely unpopulated in this snapshot. %
Second, \textbf{field verification}: when a researcher cannot be matched against the SS ground truth, field assignment is cross-checked using the top concepts from \texttt{authors.x\_concepts}.

\subsection{Perceived ethnicity inference}
\label{app:sec:inference}

Since neither Semantic Scholar nor OpenAlex provide self-reported ethnicity data for researchers, we infer \textit{perceived ethnicity} from author names---defined as the category an external observer would assign based on the name alone, consistent with prior work on algorithmic auditing of demographic bias in bibliometric data \cite{kozlowski2022avoiding}.

\para{Inference pipeline.}
We implement a two-level cascade with explicit fallback to \textit{Unknown}. At Level 1, we apply \textbf{DemographicX}\cite{liang2021demographicx}, a BERT-based transformer with combined word- and character-level embeddings that generalizes to unseen names. The model takes the full name as a single string and outputs a probability distribution over four classes: \textit{White}, \textit{Hispanic or Latino}, \textit{Black or African American}, and \textit{Asian}. If the maximum predicted probability does not exceed a confidence threshold of 0.5, inference falls back to Level 2: \textbf{EthnicolR}, an LSTM trained on the Florida voter registration dataset \cite{sood2018predicting}, whose output labels are mapped to the same four-class schema. If neither model surpasses the threshold, the record is assigned \textit{Unknown}, yielding a five-class label set: $\{$\textit{White, Asian, Black or African American, Hispanic or Latino, Unknown}$\}$. Each record retains three metadata fields: \textit{perceived\_ethnicity} which is the final class, \textit{\_\_ethnicity\_confidence} or the associated probability, and \textit{\_\_ethnicity\_source} if it is from EthnicolR or DemographicX.

To ensure consistency between the ground truth and the LLM-recommended authors compared downstream, ethnicity inference is run \textbf{once} over deduplicated researchers in the ground truth, producing a lookup artifact (\textit{researcher\_ethnicity\_lookup.csv}). Recommended authors are then assigned labels via name-based lookup against this artifact; unmatched pairs receive \textit{Unknown}.

\para{Inter-annotator agreement.}
To establish a human performance ceiling for this task, two annotators independently labeled the perceived ethnicity of 100 randomly sampled records from the classified dataset, using name alone---without access to photographs, affiliations, or publication lists. Table~\ref{tab:iaa_ethnicity} reports the results.

\begin{table}[h!]
\centering
\caption{
\textbf{Inter-annotator agreement (perceived ethnicity).}
Raw and chance-corrected agreement between annotators ($n = 100$).
}
\label{tab:iaa_ethnicity}
\begin{tabular}{lc}
\toprule
{Metric} & {Value} \\
\midrule
Raw agreement ($p$)                       & 0.64 \\
Cohen's $\kappa$                          & 0.49 \\
Krippendorff's $\alpha$ (nominal)         & 0.47 \\
\bottomrule
\end{tabular}
\end{table}

Disagreements were concentrated in the \textit{White} $\leftrightarrow$ \textit{Black or African American} pair (14 cases) and \textit{White} $\leftrightarrow$ \textit{Hispanic or Latino} (4 cases), reflecting the inherent phonological overlap between these categories in international name corpora. As a point of comparison, the same double-annotation protocol applied to an adjacent task---LLM response validity classification---yielded $\kappa = 0.83$, confirming that the lower agreement on ethnicity stems from the ambiguity of the construct itself, not from annotation error or protocol design.

\para{Limitations.}
Name-based ethnicity inference carries several well-documented limitations that apply to our pipeline. First, the inferred label is \textit{perceived}, not self-declared, and should not be interpreted as the individual's actual identity. Second, both DemographicX and EthnicolR are trained predominantly on U.S.\ data---the Florida voter registration file and U.S.\ Census records---which biases the prior distribution toward American naming patterns and may systematically misclassify researchers with non-Western names \cite{kozlowski2022avoiding}. Third, threshold-based classification is known to underestimate \textit{Black or African American} authors and overestimate \textit{White} authors relative to distributional approaches \cite{kozlowski2022avoiding}. Fourth, the four-class schema collapses substantial within-group heterogeneity---most notably within \textit{Asian}. Finally, the human ceiling of $\kappa \approx 0.49$ implies that any automatic evaluation of classifier accuracy above that value should be interpreted with caution; the task is intrinsically ambiguous even for human annotators with no time constraint.
\section{Factuality evaluation pipeline}
\label{app:sec:factuality}

Factuality is evaluated hierarchically across four dimensions: author, field, seniority, and location. Each dimension is computed by a dedicated step in a sequential pipeline.
Each step reads the output of the previous one and appends new columns; no column is overwritten, so any step can be re-run in isolation.

\paragraph{Ground-truth sources.}
Two independent bibliographic sources are used.
\emph{Semantic Scholar} (SS) provides the deduplicated researcher corpus
described in Section~D: name, primary field, inferred gender, career age ($\Delta = 2025 - \texttt{year\_first\_publication}$), and citation count. \emph{OpenAlex} (OA) supplements SS with works count, career span, and country of most recent affiliation derived from the \texttt{works} table, since \texttt{authors.last\_known\_institution} is largely unpopulated in this snapshot. The two sources are checked independently: a recommended author can be \texttt{found} in one and \texttt{not\_found} in the other, and each factuality dimension is computed from whichever source carries the relevant ground truth.

\para{Step 1: author resolution.}
For each \texttt{(name, lastname)} pair returned by an LLM, we attempt
to resolve it against both SS and OA independently.
Candidate retrieval against SS uses last-name prefix blocking (first two characters), reducing the number of pairwise comparisons from $\mathcal{O}(N \cdot M)$ to $\mathcal{O}(N \cdot M / B)$, where $B$ is the number of prefix blocks, followed by vectorized Jaro-Winkler similarity computed over the full display name. A match is accepted when display-name similarity exceeds 0.85 and every token longer than one character in the query has at least one corresponding token in the reference with similarity $\geq 0.95$ (per-token filter). The per-token filter prevents false positives that arise when partial tokens accumulate enough similarity to cross the global threshold without any single token matching well.
Against OA, resolution uses normalized exact name matching
(lower-case, accent-stripped, non-alphabetic characters collapsed),
selecting the highest-cited candidate on ties. 
Country of affiliation is derived from the most recent paper in the \texttt{works} table via chunked \texttt{ARG\_MAX} aggregation over 20 year-range partitions. %
The outcome of this step is a binary status (\textit{found} or \textit{hallucinated}) and, when an author is found, the ground-truth metadata used by subsequent steps: research field, career age, country, publication count, and citation count.

\para{Step 2: field factuality.}
For each matched author, we compare the field returned by the LLM against the ground-truth field (\texttt{gt\_field}) from Semantic Scholar, translating from Spanish and German to canonical English where necessary.

\para{Step 3: seniority factuality.}
Career age $\Delta$ is partitioned into three bins: junior ($\Delta \leq 10$), intermediate ($10 < \Delta < 20$), and senior ($\Delta \geq 20$). The LLM-requested seniority level is compared against the bin of the matched author; intermediate-age authors are counted as a mismatch for both junior and senior requests, since they
satisfy neither criterion.

\para{Step 4: location factuality.}
The location dimension uses OA exclusively, since SS does not carry
per-researcher country metadata. The LLM-specified country (one of five:
Ecuador, Germany, Japan, Canada, and South Africa, in English, Spanish,
or German) is mapped to ISO\,3166-1 alpha-2 and compared against
\texttt{oa\_country\_code}. Because this code is derived from the most
recent paper rather than a stable current affiliation, authors with
multiple affiliations or visiting positions may produce false mismatches; this is a known limitation of the OA snapshot.

\para{Manual annotation.}
To assess pipeline accuracy, two annotators independently labeled a
random sample of $n = 38$ recommendations along three binary dimensions: whether the recommended author exists (\texttt{author\_found}), whether the returned field is correct (\texttt{field\_match}), and whether the returned affiliation is correct (\texttt{affiliation\_match}). Labels were compared against the automated pipeline outputs. \Cref{tbl:factuality-validation} reports precision, recall, F1, and accuracy per dimension.

\para{Evaluation.} Author resolution achieves balanced precision and recall (both 95.2\%), indicating that the Jaro-Winkler matching with per-token filter is well calibrated in both directions. Field and affiliation checking show perfect precision but substantially lower recall, meaning the pipeline is conservative: when it predicts a match it is almost never wrong, but it misses a fraction of true matches. For field, likely causes are incomplete Spanish and German translations or sparsely populated \texttt{gt\_field} entries in SS for some authors; for affiliation, the \texttt{token\_set\_ratio} threshold of 85 and the truncated OA institution history are the primary suspects. Because accuracy is
dominated by true negatives when positives are a minority at this sample size, F$_1$ is the informative metric for these two dimensions.

\begin{table}[h]
\centering
\caption{
\textbf{Manual validation of the factuality pipeline.}
Precision, recall, F$_1$, and accuracy of each automated check, evaluated against human annotations on a sample of 38 authors ($n = 38$).
}
\label{tbl:factuality-validation}
\begin{tabular}{lrrrr}
\toprule
Dimension & Precision & Recall & F1 & Accuracy \\
\midrule
Author resolution  & 0.952 & 0.952 & 0.952 & 0.947 \\
Field match        & 1.000 & 0.500 & 0.667 & 0.947 \\
Affiliation match  & 1.000 & 0.333 & 0.500 & 0.947 \\
\bottomrule
\end{tabular}
\end{table}

\section{Results (extended)}
\begin{table}[h]
    \caption{\textbf{Residual diagnostic tests across all evaluation metrics.}
    Shapiro–Wilk statistics assess residual normality, Breusch–Pagan statistics assess heteroscedasticity, and the last column reports the proportion of fitted values outside the admissible interval (0,1) for each evaluation metric. All test statistics are statistically significant $p<0.001$.}
    \label{app:tbl:diagnostics}
    \centering
    \begin{tabular}{p{3.7cm}rrr}
    \toprule
    Metric & \makecell{Shapiro\\Wilk\\stat.} & \makecell{Breusch\\Pagan\\stat.} & \makecell{\% fitted\\values\\ $\notin [0,1]$} \\
    \midrule
    Citations parity & 0.99 & 115073.04 & 0.00 \\
    Citations middle tertile & 0.91 & 221233.80 & 0.00 \\
    Consistency & 0.79 & 105013.21 & 0.12 \\
    Works upper tertile & 0.95 & 156596.80 & 0.00 \\
    Author factuality & 0.47 & 64447.78 & 0.26 \\
    Gender parity & 0.96 & 47851.97 & 0.00 \\
    Ethinicity diversity & 0.97 & 50559.95 & 0.07 \\
    Popularity-based citations & 0.93 & 143401.56 & 0.01 \\
    Works parity & 0.98 & 117435.79 & 0.00 \\
    Validity & 0.79 & 189751.85 & 0.22 \\
    Ethinicity parity & 0.99 & 87678.47 & 0.00 \\
    Seniority factuality & 0.99 & 40547.77 & 0.16 \\
    Refusals & 0.57 & 229219.93 & 0.36 \\
    Works diversity & 0.95 & 26774.32 & 0.00 \\
    Gender diversity & 0.82 & 38265.82 & 0.00 \\
    Citations diversity & 0.96 & 28457.39 & 0.00 \\
    Popularity-based works & 0.95 & 156600.39 & 0.00 \\
    Works lower tertile & 0.96 & 152038.90 & 0.00 \\
    Citations upper tertile & 0.93 & 143398.07 & 0.01 \\
    Works middle tertile & 0.92 & 196368.95 & 0.00 \\
    Duplicates & 0.65 & 82820.31 & 0.24 \\
    Location factuality & 1.00 & 36312.63 & 0.13 \\
    Location diversity & 0.97 & 36320.18 & 0.02 \\
    Field factuality & 0.97 & 26752.33 & 0.07 \\
    Citations lower tertile & 0.96 & 139112.91 & 0.00 \\
    \bottomrule
    \end{tabular}
\end{table}

\subsection{Regression assumptions and diagnostics}
\label{app:sec:diagnostics}

Here, we report the diagnostic checks for the sensitivity models of~\Cref{sec:sensitivity}. We examine the standard regression assumptions and the behavior of the fitted values, and we show that the observed violations affect only the inferential standard errors rather than the coefficient estimates or the variance decomposition.

\para{Standard errors and significance.}
Since each prompt configuration is queried ten times, observations from the same configuration are not independent. We therefore estimate cluster-robust standard errors, clustering on the combination of prompt variables and LLM identity, with significance assessed by cluster-robust Wald tests~\cite{wald1943tests} and Benjamini--Hochberg correction~\cite{benjamini1995controlling} across all factors and metrics. 
This covariance is consistent under both heteroscedasticity and within-configuration dependence. Most effects remain significant under correction, with a small number of near-zero exceptions.

\para{Normality and homoscedasticity.}
We assess residual normality with the Shapiro--Wilk test~\cite{hanusz2016shapiro} and homoscedasticity with the Breusch--Pagan test~\cite{breusch1979simple}, per metric. \Cref{app:tbl:diagnostics} reports both statistics. At our sample size both tests reject their null hypotheses for every metric ($p < 0.001$), so the test statistics, rather than the p-values, indicate the magnitude of each departure. The Shapiro--Wilk statistic is near one for many metrics, indicating approximately normal residuals, and falls substantially for metrics such as author factuality, refusals, and duplicates. These are precisely the metrics that take a limited set of discrete values, for example the proportion of women among $k$ results. %
The departures are expected for bounded, discrete outcomes and do not indicate model misspecification.

\para{Fitted-value range.}
Because OLS does not constrain predictions to the admissible interval, we report the proportion of fitted values outside $[0,1]$ per metric in \Cref{app:tbl:diagnostics}. For most metrics this proportion is zero or negligible. It is larger for a subset of technical-quality metrics that cluster near the upper bound, reaching $0.36$ for refusals, where the linear model overshoots configurations with the highest predicted values. This is a known property of linear models applied to bounded outcomes concentrated near a limit.

\para{Consequences for inference.}
None of these conditions affects the quantities our analysis relies on. The coefficient point estimates and the sums of squares underlying $\omega^2$ are computed without distributional assumptions and are therefore unchanged. The conditions affect only the standard errors, which we address with cluster-robust covariance: this estimator is consistent under both the non-constant variance documented above and the within-configuration dependence induced by repeated sampling. Residual non-normality is immaterial at our sample size, where the sampling distribution of the coefficient estimates is approximately normal by the central limit theorem. We therefore retain OLS for the interpretability of its additive coefficients, and treat the variance decomposition and directional effects as robust to the documented violations.

\begin{figure*}[t!]
    \centering
    \includegraphics[width=\linewidth]{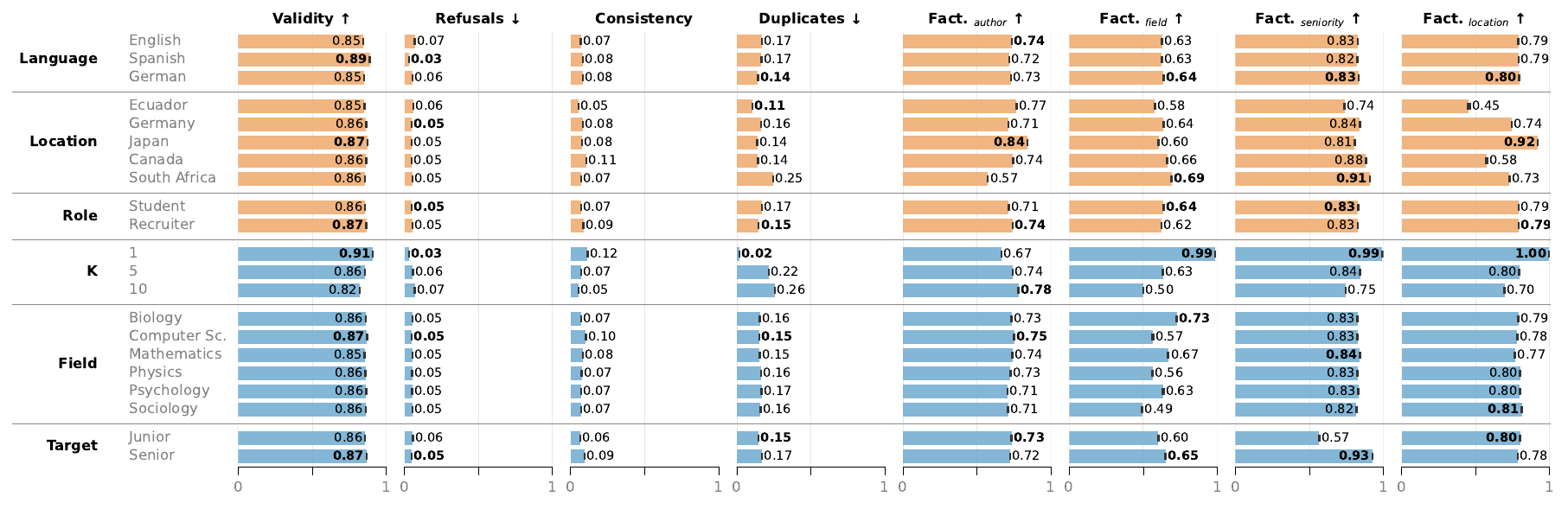}
    \caption{\textbf{Technical quality of LLM recommendations across persona and context dimensions.}
    The top-3 panels vary \emph{persona}---who is asking (language, location, role)---while the bottom-3 panels vary \emph{context}---what is asked (number of recommendations, field, target seniority). 
    Each panel reports the eight technical quality metrics for one dimension, averaged over all values of the remaining dimensions ($\uparrow$ higher is better, $\downarrow$ lower is better; \emph{Fact.} denotes factuality of the named attribute). 
    Persona dimensions (top-3 panels) yield largely stable metrics (with a few exceptions), indicating broad robustness to who the user is, whereas context dimensions (bottom-3) surface the main discrepancies in output quality.}
    \label{app:fig:rq1}
\end{figure*}
~
\begin{figure*}[t!]
    \centering
    \includegraphics[width=\linewidth]{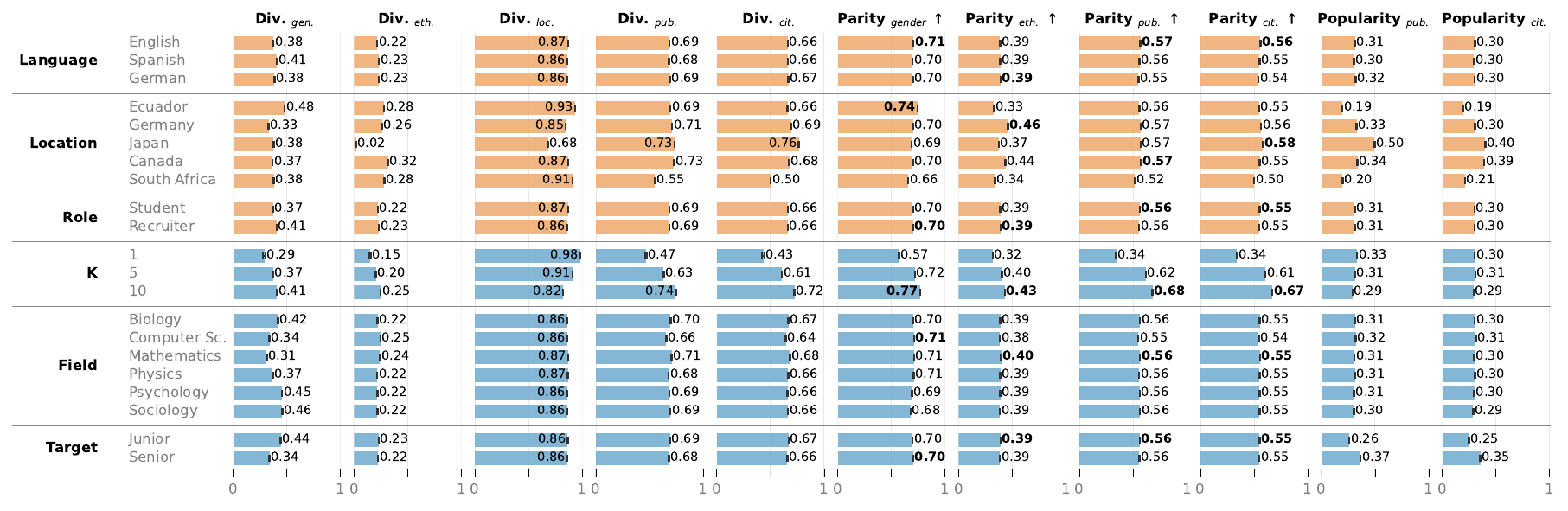}
    \caption{
    \textbf{Social representativeness of LLM recommendations across persona and context dimensions.}
    Layout follows~\Cref{app:fig:rq1}. Columns report eleven representativeness metrics grouped into three families: 
    \emph{Diversity} of the recommended set (gender, ethnicity, publications, citations), 
    \emph{Parity} between recommended and reference distributions ($\uparrow$ higher is better), and 
    \emph{Popularity} of recommended authors (publications, citations). 
    Unlike technical quality, representativeness is sensitive to both persona and context.
    }
    \label{app:fig:rq2}
\end{figure*}

\subsection{Mean metric values across prompt compositions}
\label{app:sec:mean}

\Cref{app:fig:rq1,app:fig:rq2} report the marginal mean of each metric, averaged across all values of every other prompt variable, providing a high-level view before the per-RQ breakdowns. We see that \emph{Technical quality} (\Cref{app:fig:rq1}) is largely flat across persona dimensions: validity, refusal, consistency, duplication, and factuality stay close to their global means as the language, location, and assigned role vary, with most of the spread driven by context variables ($k$, field, and target seniority). \emph{Social representativeness} (\Cref{app:fig:rq2}) behaves differently: diversity, parity with reference distributions, and the popularity of recommended scholars shift visibly under both persona and context, indicating that persona conditioning leaves a clearer fingerprint on who is surfaced than on whether the output is well-formed and factually accurate.
\subsection{Drivers of audit outcomes across persona, context, and LLM}
\label{app:sec:effects}

\begin{figure*}
    \centering
    \includegraphics[width=\linewidth]{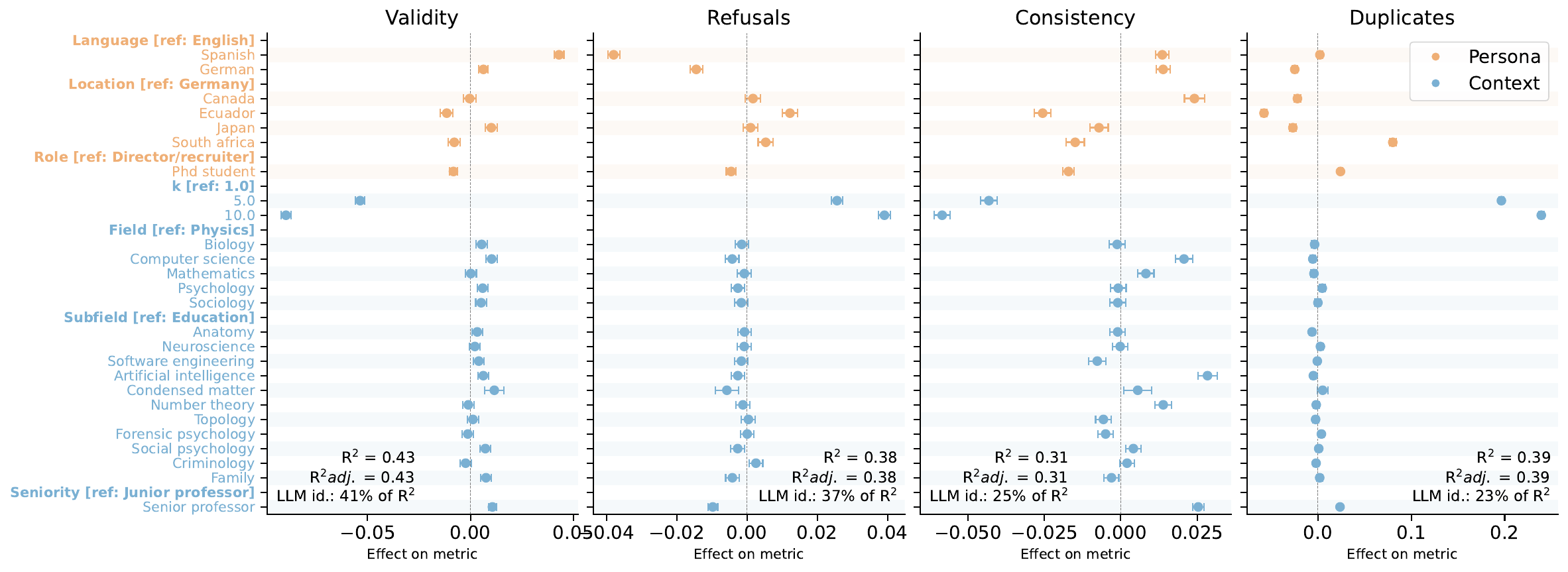}
    \caption{\textbf{Drivers of technical quality across personas, contexts, and LLMs.} Standardized regression coefficients for four technical quality outcomes (validity, refusal rate, consistency, and duplicates) as a function of persona attributes (orange: language, location, role) and context attributes (blue: $K$, field, subfield, target). Each point is the coefficient for one level of a categorical predictor relative to its reference category (shown in brackets); horizontal bars denote 95\% confidence intervals. The dashed vertical line marks zero, i.e.\ no effect relative to the reference. Each panel reports $R^2$, adjusted $R^2$, and the share of explained variance attributable to LLM identity (included as fixed effects). 
    Three patterns stand out. First, $K$ is the dominant context-side lever: increasing the number of recommendations per prompt lowers validity and consistency while raising refusals and duplicates, uniformly degrading technical quality. 
    Second, persona attributes, particularly prompt language and location, shift all four metrics by comparable magnitudes. Field and subfield coefficients, by contrast, hover near zero across all four metrics. 
    Third, LLM identity explains a large share of variance in validity (41\%) and refusals (37\%), a moderate share for consistency (23\%), and only a small share for duplicates (6\%), indicating that model choice governs content-level reliability while duplicates is driven primarily by prompt parameters.}
    \label{app:fig:me:technical}
\end{figure*}

\begin{figure*}
    \centering
    \includegraphics[width=\linewidth]{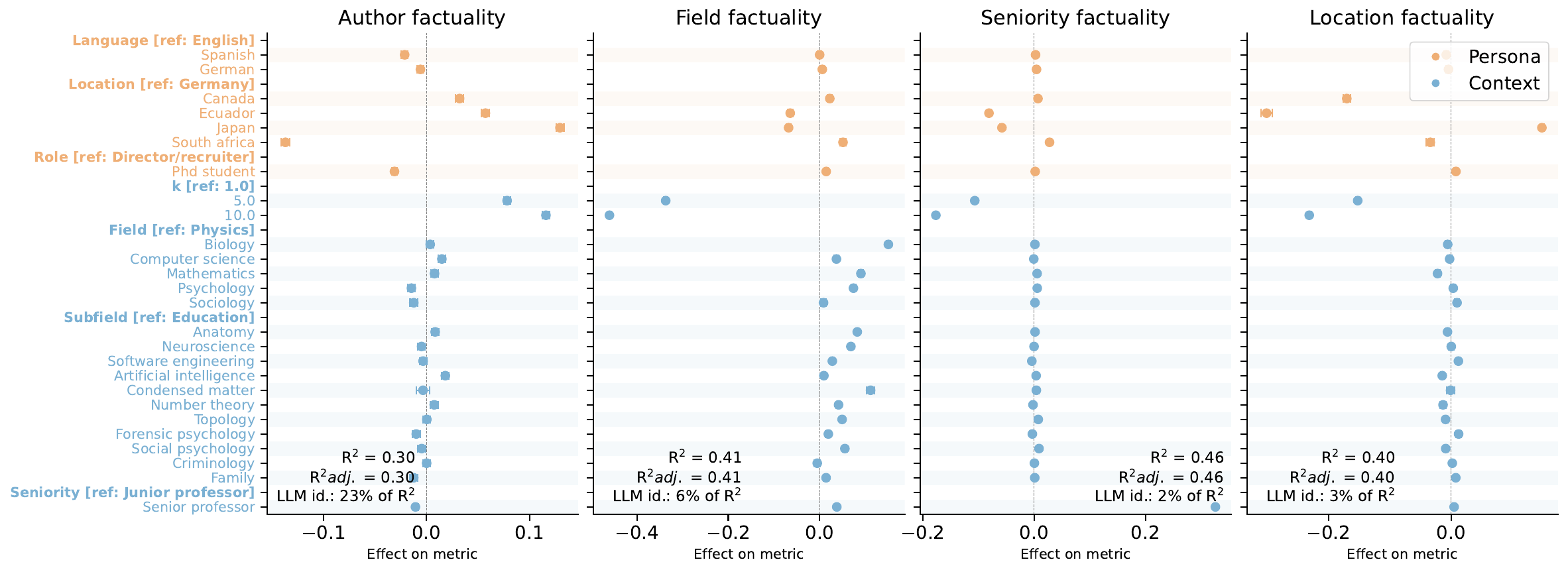}
    \caption{\textbf{Drivers of factuality across personas, context, and LLMs.} 
    Same specification as \Cref{app:fig:me:technical}, applied to four factuality outcomes: author, field, seniority, and location. Three patterns stand out. 
    First, $K$ remains the dominant context-side lever, with larger values sharply degrading factuality of field, seniority, and location. 
    Second, persona attributes exert their strongest influence on the factuality dimension they directly correspond to: persona location, for instance, drives factuality location far more than the other outcomes, with Japan boosting accuracy and Ecuador, Canada, and South Africa reducing it. 
    Third, in striking contrast to the other technical quality variables in~\Cref{app:fig:me:technical}, LLM identity explains only a marginal share of variance across all four factuality outcomes, indicating that factual accuracy is governed predominantly by prompt and context features rather than by model choice. 
    Finally, factuality of author is markedly harder to predict than the other three outcomes, with the model explaining far less of its variance.
    }
    \label{app:fig:me:factuality}
\end{figure*}

\begin{figure*}
    \centering
    \includegraphics[width=\linewidth]{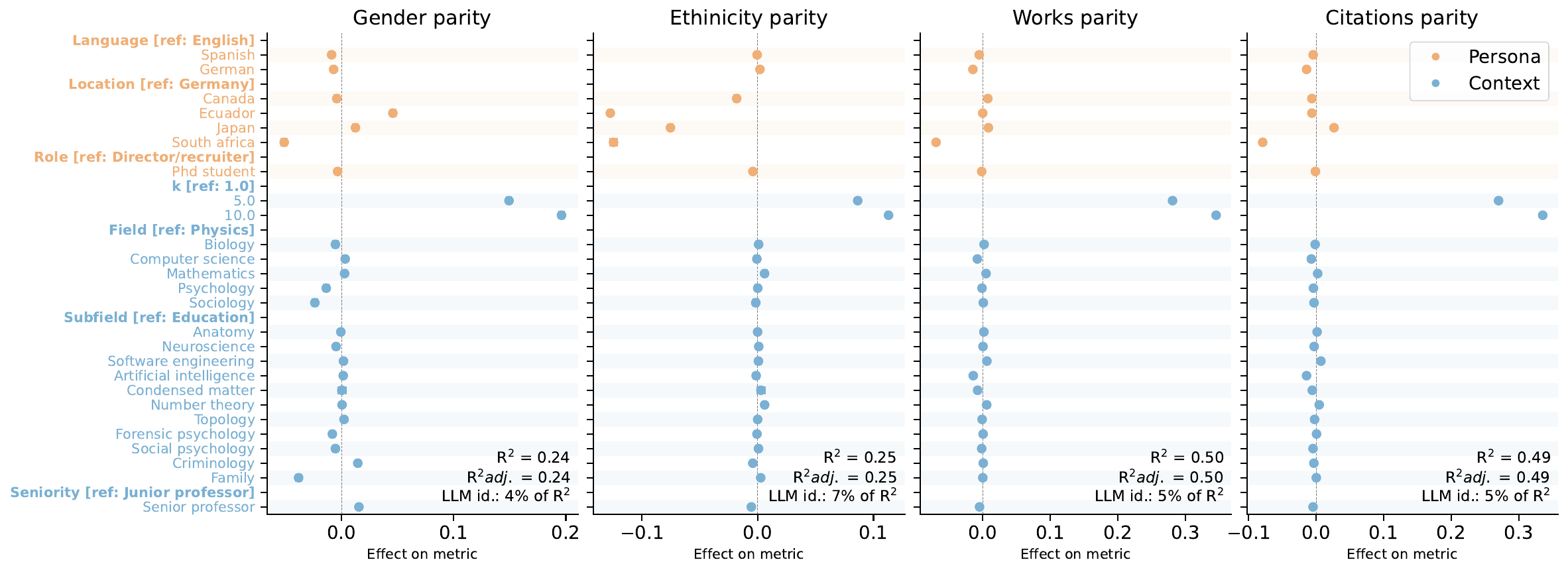}
    \caption{\textbf{Drivers of parity across personas, context, and LLMs.} 
    Same specification as \Cref{app:fig:me:technical}, applied to four parity outcomes: gender, ethnicity, works, and citations. Three patterns stand out. 
    First, $K$ overwhelmingly dominates all other predictors, with larger values substantially increasing parity across every outcome, consistent with the mechanical effect of drawing more recommendations per prompt. 
    Second, persona location is the most influential persona attribute, with South Africa standing out as consistently reducing parity across all four outcomes. Third, as with factuality (\Cref{app:fig:me:factuality}), LLM identity explains only a small share of variance. %
    }
    \label{app:fig:me:parity}
\end{figure*}

\begin{figure*}
    \centering
    \includegraphics[width=\linewidth]{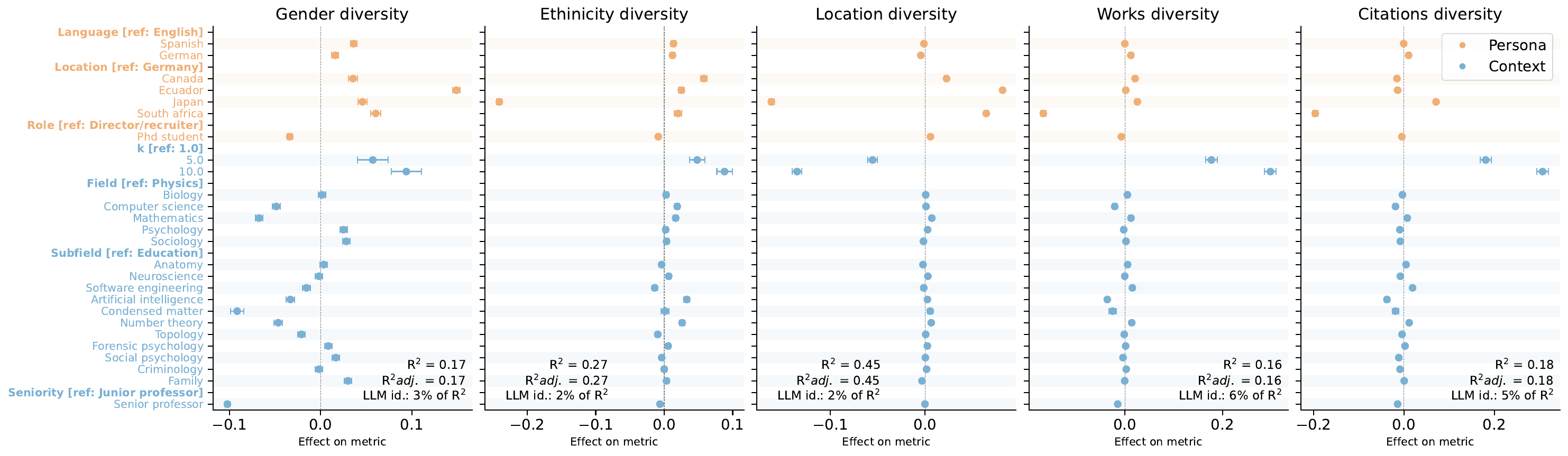}
    \caption{\textbf{Drivers of diversity across personas, context, and LLMs.} 
    Same specification as \Cref{app:fig:me:technical}, applied to three diversity outcomes: gender, ethnicity, and location. 
    Three patterns stand out. 
    First, $K$ acts in opposite directions depending on the outcome, increasing gender and ethnicity diversity while reducing location diversity, suggesting that drawing more generations expands the demographic pool but concentrates geographic responses. 
    Second, persona location exerts strong and asymmetric effects, with Japan markedly reducing diversity across all three outcomes and Ecuador increasing it. 
    Third, field and subfield effects appear almost exclusively in gender diversity, with male-dominated fields such as computer science and mathematics increasing gender diversity and female-leaning ones such as sociology and biology decreasing it, while ethnicity and location diversity remain insensitive to field and subfield. Finally, targeting a senior professor substantially lowers gender diversity relative to a junior professor, consistent with stronger gender skew in recommendations for more established roles. 
    LLM identity once again contributes negligibly to explained variance. %
    }
    \label{app:fig:me:diversity}
\end{figure*}

\begin{figure*}
    \centering
    \includegraphics[width=\linewidth]{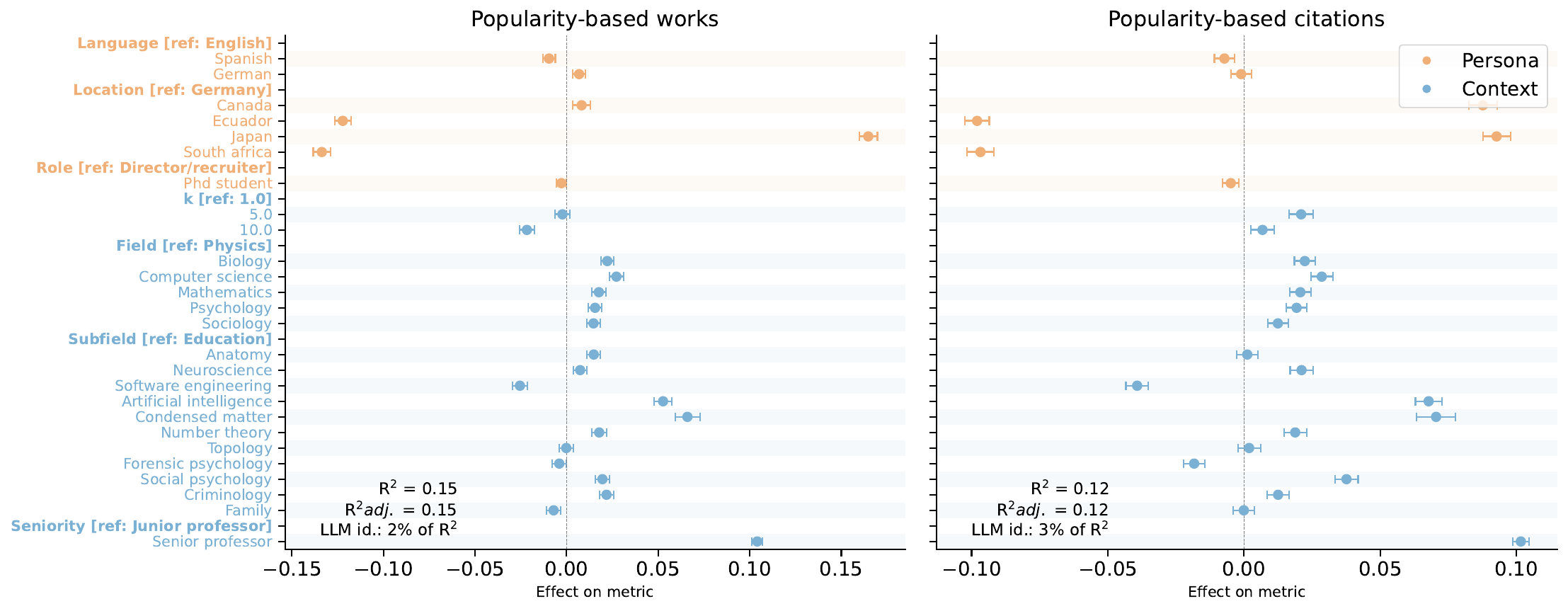}
    \caption{\textbf{Drivers of popularity across personas, context, and LLMs.} Same specification as~\Cref{app:fig:me:technical}, applied to two popularity outcomes: the share of recommended authors falling in the upper tertile of the corpus-wide distribution of works (left) and citations (right). 
    Three patterns stand out. First, persona \texttt{location}~is the dominant driver: a Japan-based persona sharply increases the share of upper-tertile recommendations for both works and citations, while South Africa and Ecuador based personas shift recommendations away from the upper tertile. 
    Second, in contrast to other outcomes, \texttt{k}~plays only a marginal role. 
    Third, targeting a senior professor predictably increases the share of upper-tertile recommendations.
    Overall explanatory power is modest ($R^2 = 0.15$ for works, $0.12$ for citations), with LLM identity accounting for only a small share (2--3\% of $R^2$), leaving substantial variance unexplained by the measured persona and context attributes.
    }
    \label{app:fig:me:popularity}
\end{figure*}

\Cref{app:fig:me:technical,app:fig:me:factuality,app:fig:me:parity,app:fig:me:diversity,app:fig:me:popularity} report regression coefficients from the fixed-effects specification described in~\Cref{sec:methods}, applied in turn to each outcome family: technical quality, factuality, parity, diversity, and popularity. Persona attributes (language, location, role) and context attributes ($K$, field, subfield, target seniority) enter as categorical predictors relative to their reference levels, with LLM identity absorbed as fixed effects. Each panel reports $R^2$, adjusted $R^2$, and the share of $R^2$ attributable to LLM identity, which separates the contribution of model choice from that of prompt composition.

\subsection{By model}
\label{app:sec:models}

\begin{figure*}[ht]
    \centering
    \includegraphics[width=\linewidth]{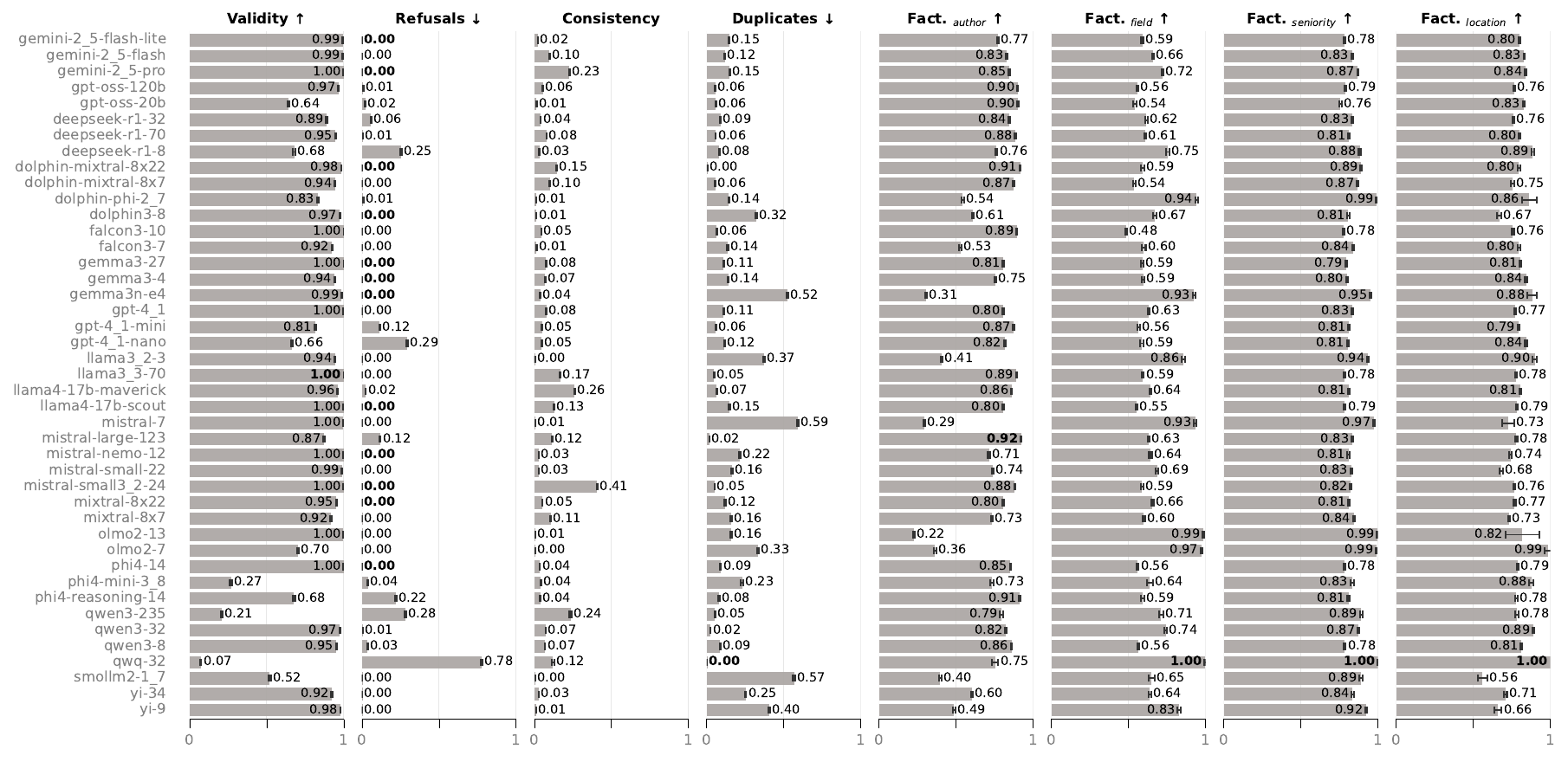}
    \caption{\textbf{Technical quality by model.} 
    Mean score on each metric (columns) for every LLM in our sample (rows). Arrows indicate the desirable direction ($\uparrow$ higher is better, $\downarrow$ lower is better). Three patterns stand out. 
    First, validity and refusal behaviour cleanly separate models: most produce valid, non-refusing responses nearly all the time, while a small set of outliers (e.g., qwq-32, qwen3-235, phi4-mini-3.8) fail on validity or refuse a substantial share of prompts. 
    Second, factuality of author is uniformly high across nearly all models, whereas factuality of field, seniority, and location is markedly lower and far more variable, indicating that models can name plausible researchers but often misplace them along these dimensions. 
    Third, the few models with very high factuality scores on these harder dimensions (e.g., qwq-32, olmo2-13, deepseek-r1-8) are typically those with low validity or high refusal rates, suggesting that their elevated scores reflect selection on a small set of completed responses rather than genuinely better factual grounding.}
    \label{app:fig:model:tech}
\end{figure*}

\begin{figure*}[ht]
    \centering
    \includegraphics[width=\linewidth]{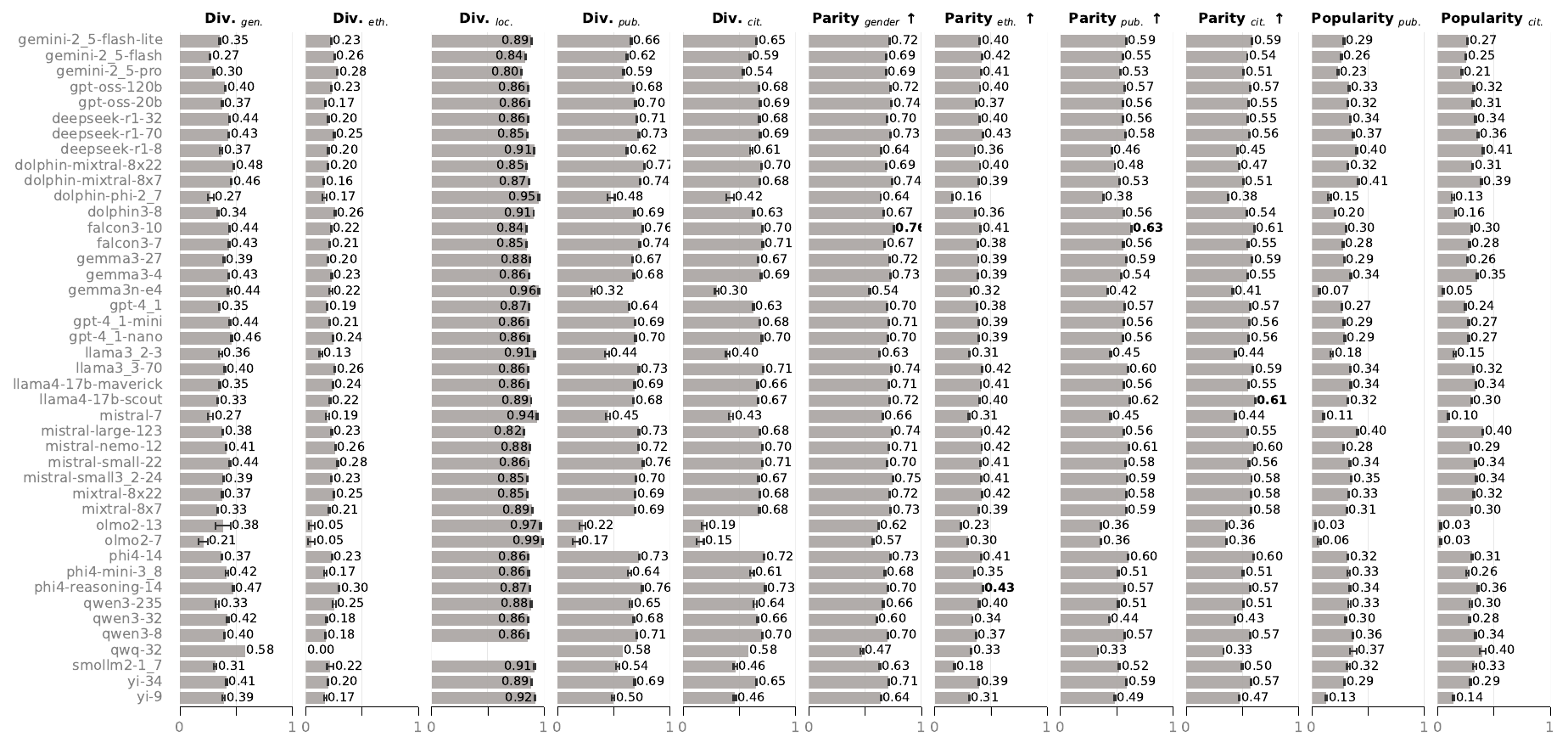}
    \caption{\textbf{Social representativeness by model.} Mean score on each metric (columns) for every LLM in our sample (rows). Arrows indicate the desirable direction ($\uparrow$ higher is better). Diversity scores are broadly comparable across models, while parity and popularity metrics show substantially more variation, with no single model dominating across all dimensions.}
    \label{app:fig:model:social}
\end{figure*}

\begin{figure*}[ht]
    \centering
    \includegraphics[width=\linewidth]{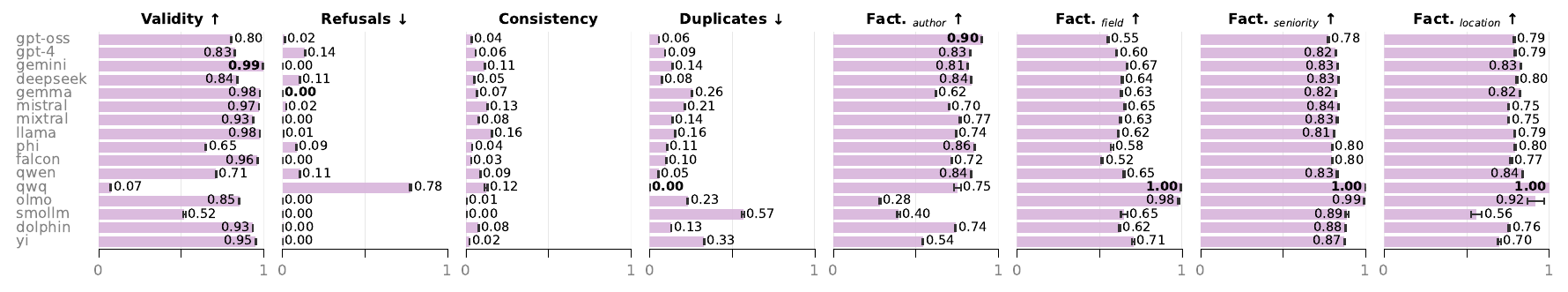}
    \caption{\textbf{Technical quality and factuality by model family.} 
    Mean score on each metric (columns), aggregated across all models within each family (rows). Arrows indicate the desirable direction ($\uparrow$ higher is better, $\downarrow$ lower is better). Same metrics as \Cref{app:fig:model:tech}, collapsed to the family level.}
    \label{app:fig:modelfam:tech}
\end{figure*}

\begin{figure*}[ht]
    \centering
    \includegraphics[width=\linewidth]{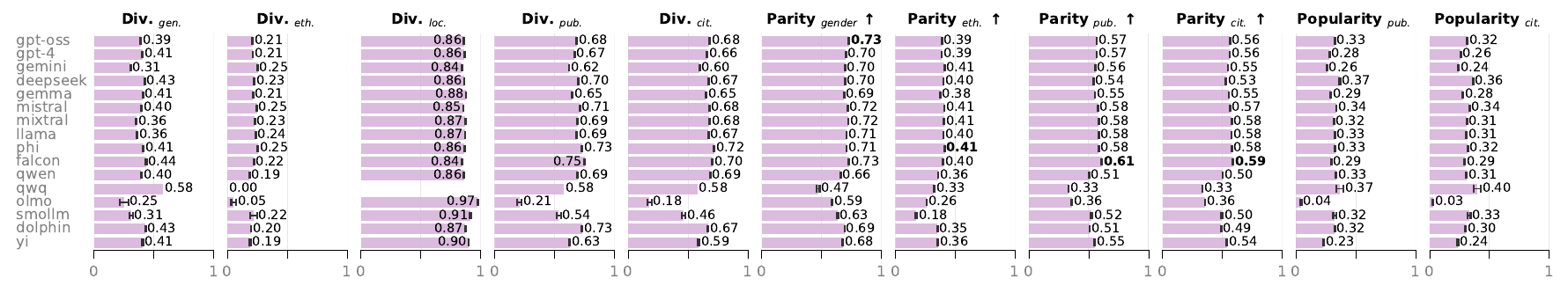}
    \caption{\textbf{Social representativeness by model family.} Mean score on each metric (columns), aggregated across all models within each family (rows). Arrows indicate the desirable direction ($\uparrow$ higher is better). Same metrics as \Cref{app:fig:model:social}, collapsed to the family level.
    }
    \label{app:fig:modelfam:social}
\end{figure*}

\begin{figure*}[ht]
    \centering
    \begin{subfigure}[t]{\textwidth}
        \centering
        \includegraphics[width=1.0\textwidth]{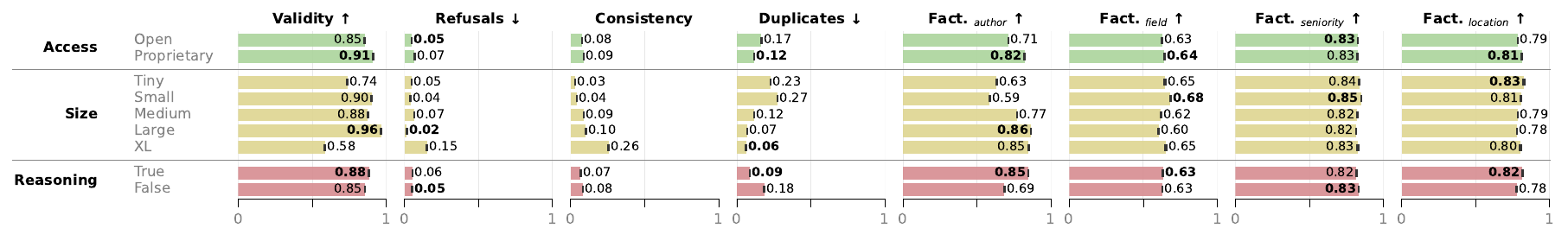}
        \caption{Technical quality}
    \end{subfigure}
    \begin{subfigure}[t]{\textwidth}
        \centering
        \includegraphics[width=1.0\textwidth]{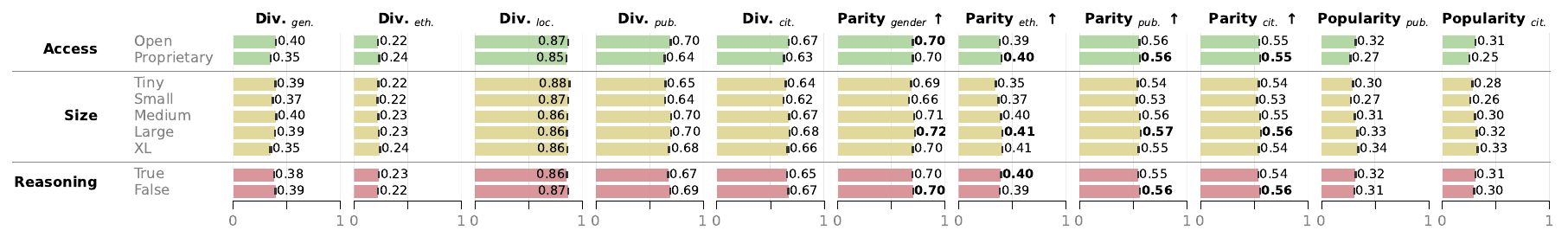}
        \caption{Social Representativeness}
    \end{subfigure}
    \caption{\textbf{Performance of audited LLMs across infrastructural characteristics.}
    Performance comparison of the audited LLMs grouped by three infrastructural characteristics: access type (open-weight vs. proprietary), model size, and reasoning capability. 
    Panel (a) reports technical quality metrics, including validity, refusals, consistency, duplicates, and factuality dimensions (author, field, seniority, and location). 
    Panel (b) reports social representativeness metrics, including diversity (Div.), parity, and popularity across demographic and publication attributes. 
    Larger models generally achieve higher validity and factuality scores, although XL models also exhibit higher refusal rates and greater consistency variation. 
    Proprietary models slightly outperform open-weight models on validity and author and location factuality, while differences in social representativeness remain limited. 
    Reasoning-enabled models improve several factuality and parity metrics, but at the cost of increased refusals. 
    }
    \label{fig:performance_main}
\end{figure*}

\Cref{app:fig:model:tech,app:fig:model:social} report mean scores per metric for each of the 43~LLMs. \Cref{app:fig:modelfam:tech,app:fig:modelfam:social} report the same metrics by model family. \Cref{fig:performance_main} shows models grouped by three infrastructural characteristics: access type (open-weight versus proprietary), parameter size, and reasoning capability. Together they provide a descriptive view of how outcomes vary across models, families, and infrastructure profiles.

\begin{thebibliography}{51}
\providecommand{\natexlab}[1]{#1}

\bibitem[{Abdin et~al.(2024)Abdin, Gunasekar, Chandrasekaran, Li, Yuksekgonul,
  Peshawaria, Naik, and Nushi}]{abdin2024kitab}
Abdin, M.~I.; Gunasekar, S.; Chandrasekaran, V.; Li, J.; Yuksekgonul, M.;
  Peshawaria, R.; Naik, R.; and Nushi, B. 2024.
\newblock Kitab: Evaluating llms on constraint satisfaction for information
  retrieval.
\newblock In \emph{International Conference on Learning Representations},
  volume 2024, 30664--30686.

\bibitem[{Anonymous(2025)}]{anonymous2025repo}
Anonymous. 2025.
\newblock Anonymous Repository for Auditing LLMs as People Recommender Systems
  Across Languages and Countries.
\newblock \url{https://anonymous.4open.science/r/PersonasScholarRec}.
\newblock Anonymous repository for double-blind review.

\bibitem[{Anzenberg et~al.(2025)Anzenberg, Samajpati, Chandrasekar, and
  Kacholia}]{anzenberg2025evaluating}
Anzenberg, E.; Samajpati, A.; Chandrasekar, S.; and Kacholia, V. 2025.
\newblock Evaluating the Promise and Pitfalls of LLMs in Hiring Decisions.
\newblock \emph{arXiv preprint arXiv:2507.02087}.

\bibitem[{Awasthi, Rao, and Jayagopi(2025)}]{awasthi2025resumegenai}
Awasthi, D.; Rao, P.~S.; and Jayagopi, D.~B. 2025.
\newblock ResumeGenAI: Supporting Job Seekers with LLM-Driven Resume Feedback.
\newblock In \emph{Proceedings of the 7th ACM Conference on Conversational User
  Interfaces}, 1--9.

\bibitem[{Barolo et~al.(2025)Barolo, Valentin, Karimi, Gal{\'a}rraga,
  M{\'e}ndez, and Esp{\'\i}n-Noboa}]{barolo2025whose}
Barolo, D.; Valentin, C.; Karimi, F.; Gal{\'a}rraga, L.; M{\'e}ndez, G.~G.; and
  Esp{\'\i}n-Noboa, L. 2025.
\newblock Whose Name Comes Up? Auditing LLM-Based Scholar Recommendations.
\newblock \emph{arXiv preprint arXiv:2506.00074}.

\bibitem[{Benjamini and Hochberg(1995)}]{benjamini1995controlling}
Benjamini, Y.; and Hochberg, Y. 1995.
\newblock Controlling the false discovery rate: a practical and powerful
  approach to multiple testing.
\newblock \emph{Journal of the Royal statistical society: series B
  (Methodological)}, 57(1): 289--300.

\bibitem[{Breusch and Pagan(1979)}]{breusch1979simple}
Breusch, T.~S.; and Pagan, A.~R. 1979.
\newblock A simple test for heteroscedasticity and random coefficient
  variation.
\newblock \emph{Econometrica: Journal of the econometric society}, 1287--1294.

\bibitem[{Cheng et~al.(2024)Cheng, Edara, Zhang, Kejriwal, and
  Calyam}]{cheng2024influence}
Cheng, X.; Edara, L.~S.; Zhang, Y.; Kejriwal, M.; and Calyam, P. 2024.
\newblock Influence Role Recognition and LLM-Based Scholar Recommendation in
  Academic Social Networks.
\newblock In \emph{2024 IEEE 11th International Conference on Data Science and
  Advanced Analytics (DSAA)}, 1--11. IEEE.

\bibitem[{De~Araujo et~al.(2025)De~Araujo, R{\"o}ttger, Hovy, and
  Roth}]{de2025principled}
De~Araujo, P. H.~L.; R{\"o}ttger, P.; Hovy, D.; and Roth, B. 2025.
\newblock Principled personas: Defining and measuring the intended effects of
  persona prompting on task performance.
\newblock In \emph{Proceedings of the 2025 Conference on Empirical Methods in
  Natural Language Processing}, 26845--26874.

\bibitem[{Durmus et~al.(2023)Durmus, Nguyen, Liao, Schiefer, Askell, Bakhtin,
  Chen, Hatfield-Dodds, Hernandez, Joseph et~al.}]{durmus2023towards}
Durmus, E.; Nguyen, K.; Liao, T.~I.; Schiefer, N.; Askell, A.; Bakhtin, A.;
  Chen, C.; Hatfield-Dodds, Z.; Hernandez, D.; Joseph, N.; et~al. 2023.
\newblock Towards measuring the representation of subjective global opinions in
  language models.
\newblock \emph{arXiv preprint arXiv:2306.16388}.

\bibitem[{Espin-Noboa and Mendez(2026)}]{espin2026whose}
Espin-Noboa, L.; and Mendez, G.~G. 2026.
\newblock Whose Name Comes Up? Benchmarking and Intervention-Based Auditing of
  LLM-Based Scholar Recommendation.
\newblock \emph{arXiv preprint arXiv:2602.08873}.

\bibitem[{Fabbri et~al.(2022)Fabbri, Croci, Bonchi, and
  Castillo}]{fabbri2022exposure}
Fabbri, F.; Croci, M.~L.; Bonchi, F.; and Castillo, C. 2022.
\newblock Exposure inequality in people recommender systems: The long-term
  effects.
\newblock In \emph{Proceedings of the international AAAI conference on web and
  social media}, volume~16, 194--204.

\bibitem[{Hanusz, Tarasinska, and Zielinski(2016)}]{hanusz2016shapiro}
Hanusz, Z.; Tarasinska, J.; and Zielinski, W. 2016.
\newblock Shapiro--Wilk test with known mean.
\newblock \emph{REVSTAT-statistical Journal}, 14(1): 89--100.

\bibitem[{Hu and Collier(2024)}]{hu2024personaeffect}
Hu, T.; and Collier, N. 2024.
\newblock Quantifying the Persona Effect in LLM Simulations.
\newblock \emph{arXiv preprint arXiv:2402.10811}.

\bibitem[{Jaramillo et~al.(2025)Jaramillo, Macedo, Oliveira, Karimi, and
  Menezes}]{jaramillo2025systematic}
Jaramillo, A.~M.; Macedo, M.; Oliveira, M.; Karimi, F.; and Menezes, R. 2025.
\newblock Systematic comparison of gender inequality in scientific rankings
  across disciplines.
\newblock \emph{arXiv preprint arXiv:2501.13061}.

\bibitem[{Jiao et~al.(2025)Jiao, Afroogh, Xu, and
  Phillips}]{jiao2025navigating}
Jiao, J.; Afroogh, S.; Xu, Y.; and Phillips, C. 2025.
\newblock Navigating llm ethics: Advancements, challenges, and future
  directions.
\newblock \emph{AI and Ethics}, 1--25.

\bibitem[{Karimi et~al.(2016)Karimi, Wagner, Lemmerich, Jadidi, and
  Strohmaier}]{karimi2016inferring}
Karimi, F.; Wagner, C.; Lemmerich, F.; Jadidi, M.; and Strohmaier, M. 2016.
\newblock Inferring Gender from Names on the Web: A Comparative Evaluation of
  Gender Detection Methods.
\newblock In \emph{Proceedings of the 25th International Conference Companion
  on World Wide Web}, WWW '16 Companion, 53--54. New York, New York, USA: ACM
  Press.

\bibitem[{Kim, Yang, and Jung(2025)}]{kim2025persona}
Kim, J.; Yang, N.; and Jung, K. 2025.
\newblock Persona is a Double-Edged Sword: Rethinking the Impact of Role-play
  Prompts in Zero-shot Reasoning Tasks.
\newblock In \emph{Proceedings of the 14th International Joint Conference on
  Natural Language Processing and the 4th Conference of the Asia-Pacific
  Chapter of the Association for Computational Linguistics}, 848--862.

\bibitem[{Kinney et~al.(2023)Kinney, Anastasiades, Authur, Beltagy, Bragg,
  Buraczynski, Cachola, Candra, Chandrasekhar, Cohan, Crawford, Downey,
  Dunkelberger, Etzioni, Evans, Feldman, Gorney, Graham, Hu, Huff, King,
  Kohlmeier, Kuehl, Langan, Lin, Liu, Lo, Lochner, MacMillan, Murray, Newell,
  Rao, Rohatgi, Sayre, Shen, Singh, Soldaini, Subramanian, Tanaka, Wade,
  Wagner, Wang, Wilhelm, Wu, Yang, Zamarron, van Zuylen, and
  Weld}]{Kinney2023TheSS}
Kinney, R.~M.; Anastasiades, C.; Authur, R.; Beltagy, I.; Bragg, J.;
  Buraczynski, A.; Cachola, I.; Candra, S.; Chandrasekhar, Y.; Cohan, A.;
  Crawford, M.; Downey, D.; Dunkelberger, J.; Etzioni, O.; Evans, R.; Feldman,
  S.; Gorney, J.; Graham, D.~W.; Hu, F.; Huff, R.; King, D.; Kohlmeier, S.;
  Kuehl, B.; Langan, M.; Lin, D.; Liu, H.; Lo, K.; Lochner, J.; MacMillan, K.;
  Murray, T.~C.; Newell, C.; Rao, S.~R.; Rohatgi, S.; Sayre, P.; Shen, S.~Z.;
  Singh, A.; Soldaini, L.; Subramanian, S.; Tanaka, A.; Wade, A.~D.; Wagner,
  L.~M.; Wang, L.~L.; Wilhelm, C.; Wu, C.; Yang, J.; Zamarron, A.; van Zuylen,
  M.; and Weld, D.~S. 2023.
\newblock The Semantic Scholar Open Data Platform.
\newblock \emph{ArXiv}, abs/2301.10140.

\bibitem[{Kozlowski et~al.(2022)Kozlowski, Murray, Bell, Hulsey, Larivière,
  Monroe-White, and Sugimoto}]{kozlowski2022avoiding}
Kozlowski, D.; Murray, D.~S.; Bell, A.; Hulsey, W.; Larivière, V.;
  Monroe-White, T.; and Sugimoto, C.~R. 2022.
\newblock Avoiding bias when inferring race using name-based approaches.
\newblock \emph{PLOS ONE}.

\bibitem[{Kroes and Finley(2023)}]{kroes2023demystifying}
Kroes, A.~D.; and Finley, J.~R. 2023.
\newblock Demystifying omega squared: Practical guidance for effect size in
  common analysis of variance designs.
\newblock \emph{Psychological Methods}.

\bibitem[{Landis and Koch(1977)}]{landis1977measurement}
Landis, J.~R.; and Koch, G.~G. 1977.
\newblock The measurement of observer agreement for categorical data.
\newblock \emph{biometrics}, 159--174.

\bibitem[{Letteri and Vittorini(2024)}]{letteri2024exploring}
Letteri, I.; and Vittorini, P. 2024.
\newblock Exploring the Impact of LLM-Generated Feedback: Evaluation from
  Professors and Students in Data Science Courses.
\newblock In \emph{International Conference in Methodologies and intelligent
  Systems for Techhnology Enhanced Learning}, 11--20. Springer.

\bibitem[{Li, Qin, and Sheng(2025)}]{li2025multi}
Li, T.; Qin, Y.; and Sheng, O. R.~L. 2025.
\newblock A Multi-Task Evaluation of LLMs' Processing of Academic Text Input.
\newblock \emph{arXiv preprint arXiv:2508.11779}.

\bibitem[{Liang and Acuna(2021)}]{liang2021demographicx}
Liang, L.; and Acuna, D. 2021.
\newblock demographicx: A Python package for estimating gender and ethnicity
  using deep learning transformers.
\newblock \emph{Zenodo https://doi. org/10.5281/zenodo}, 4898367.

\bibitem[{Lin et~al.(2022)Lin, Wang, Zhu, and Caverlee}]{lin2022quantifying}
Lin, A.; Wang, J.; Zhu, Z.; and Caverlee, J. 2022.
\newblock Quantifying and mitigating popularity bias in conversational
  recommender systems.
\newblock In \emph{Proceedings of the 31st ACM international conference on
  information \& knowledge management}, 1238--1247.

\bibitem[{Liu et~al.(2025)Liu, Elekes, Lu, Dorantes-Gilardi, and
  Barab{\'a}si}]{liu2025unequal}
Liu, Y.; Elekes, {\'A}.; Lu, J.; Dorantes-Gilardi, R.; and Barab{\'a}si, A.-L.
  2025.
\newblock Unequal Scientific Recognition in the Age of LLMs.
\newblock In \emph{Findings of the Association for Computational Linguistics:
  EMNLP 2025}, 23558--23568.

\bibitem[{Lo et~al.(2025)Lo, Qiu, Wang, Yu, Chen, Zhang, and Lo}]{Lo_2025_CVPR}
Lo, F. P.-W.; Qiu, J.; Wang, Z.; Yu, H.; Chen, Y.; Zhang, G.; and Lo, B. 2025.
\newblock AI Hiring with LLMs: A Context-Aware and Explainable Multi-Agent
  Framework for Resume Screening.
\newblock In \emph{Proceedings of the IEEE/CVF Conference on Computer Vision
  and Pattern Recognition (CVPR) Workshops}, 4223--4232.

\bibitem[{Lutz et~al.(2025)Lutz, Sen, Ahnert, Rogers, and
  Strohmaier}]{lutz2025prompt}
Lutz, M.; Sen, I.; Ahnert, G.; Rogers, E.; and Strohmaier, M. 2025.
\newblock The Prompt Makes the Person(a): A Systematic Evaluation of
  Sociodemographic Persona Prompting for Large Language Models.
\newblock In \emph{Findings of the Association for Computational Linguistics:
  EMNLP 2025}, 23212--23237. Suzhou, China: Association for Computational
  Linguistics.

\bibitem[{Pava et~al.(2025)Pava, Meinhardt, Zaman, Friedman, Truong, Zhang,
  Marivate, and Koyejo}]{pava2025mind}
Pava, J.; Meinhardt, C.; Zaman, H. B.~U.; Friedman, T.; Truong, S.~T.; Zhang,
  D.; Marivate, V.; and Koyejo, S. 2025.
\newblock Mind the (Language) Gap: Mapping the Challenges of LLM Development in
  Low-Resource Language Contexts.

\bibitem[{Polonioli(2021)}]{polonioli2021ethics}
Polonioli, A. 2021.
\newblock The ethics of scientific recommender systems.
\newblock \emph{Scientometrics}, 126(2): 1841--1848.

\bibitem[{Priem, Piwowar, and Orr(2022)}]{priem2022openalex}
Priem, J.; Piwowar, H.; and Orr, R. 2022.
\newblock OpenAlex: A fully-open index of scholarly works, authors, venues,
  institutions, and concepts.
\newblock \emph{arXiv preprint arXiv:2205.01833}.

\bibitem[{Sakib and Bijoy~Das(2024)}]{sakib2024challengingfairness}
Sakib, S.~K.; and Bijoy~Das, A. 2024.
\newblock Challenging Fairness: A Comprehensive Exploration of Bias in
  LLM-Based Recommendations.
\newblock In \emph{2024 IEEE International Conference on Big Data (BigData)},
  1585--1592.

\bibitem[{Sandnes(2024)}]{sandnes2024can}
Sandnes, F.~E. 2024.
\newblock Can we identify prominent scholars using ChatGPT?
\newblock \emph{Scientometrics}, 129(1): 713--718.

\bibitem[{Sood and Laohaprapanon(2018)}]{sood2018predicting}
Sood, G.; and Laohaprapanon, S. 2018.
\newblock Predicting Race and Ethnicity From the Sequence of Characters in a
  Name.
\newblock \emph{arXiv preprint arXiv:1805.02109}.

\bibitem[{Tonneau et~al.(2026)Tonneau, Sehgal, Malhotra, Kazemi, Orozco-Olvera,
  Mu{\~n}oz~Boudet, Subramanian, Fraiberger, Guntuku, and
  Hofmann}]{tonneau2026demographic}
Tonneau, M.; Sehgal, N. K.~R.; Malhotra, N.; Kazemi, S.; Orozco-Olvera, V.;
  Mu{\~n}oz~Boudet, A.~M.; Subramanian, L.; Fraiberger, S.~P.; Guntuku, S.~C.;
  and Hofmann, V. 2026.
\newblock Different {Demographic} {Cues} {Yield} {Inconsistent} {Conclusions}
  {About} {LLM} {Personalization} and {Bias}.
\newblock \emph{arXiv preprint arXiv:2601.18486v2}.

\bibitem[{V{\'a}s{\'a}rhelyi and Horv{\'a}t(2023)}]{vasarhelyi2023benefits}
V{\'a}s{\'a}rhelyi, O.; and Horv{\'a}t, E.-{\'A}. 2023.
\newblock Who benefits from altmetrics? The effect of team gender composition
  on the link between online visibility and citation impact.
\newblock \emph{arXiv preprint arXiv:2308.00405}.

\bibitem[{V{\'a}s{\'a}rhelyi et~al.(2021)V{\'a}s{\'a}rhelyi, Zakhlebin,
  Milojevi{\'c}, and Horv{\'a}t}]{vasarhelyi2021gender}
V{\'a}s{\'a}rhelyi, O.; Zakhlebin, I.; Milojevi{\'c}, S.; and Horv{\'a}t,
  E.-{\'A}. 2021.
\newblock Gender inequities in the online dissemination of scholars’ work.
\newblock \emph{Proceedings of the National Academy of Sciences}, 118(39):
  e2102945118.

\bibitem[{Wald(1943)}]{wald1943tests}
Wald, A. 1943.
\newblock Tests of statistical hypotheses concerning several parameters when
  the number of observations is large.
\newblock \emph{Transactions of the American Mathematical society}, 54(3):
  426--482.

\bibitem[{Wang, Ho, and Koyejo(2025)}]{wang2025inadequacy}
Wang, A.; Ho, D.~E.; and Koyejo, S. 2025.
\newblock The inadequacy of offline large language model evaluations: A need to
  account for personalization in model behavior.
\newblock \emph{Patterns}, 6(12).

\bibitem[{Wang et~al.(2024)Wang, Wang, Manzoor, Liu, Georgiev, Das, and
  Nakov}]{wang-etal-2024-factuality}
Wang, Y.; Wang, M.; Manzoor, M.~A.; Liu, F.; Georgiev, G.~N.; Das, R.~J.; and
  Nakov, P. 2024.
\newblock Factuality of Large Language Models: A Survey.
\newblock In Al-Onaizan, Y.; Bansal, M.; and Chen, Y.-N., eds.,
  \emph{Proceedings of the 2024 Conference on Empirical Methods in Natural
  Language Processing}, 19519--19529. Miami, Florida, USA: Association for
  Computational Linguistics.

\bibitem[{Weeber et~al.(2026)Weeber, Neplenbroek, Batzner, and
  Pad{\'o}}]{weeber2026persona}
Weeber, F.; Neplenbroek, V.; Batzner, J.; and Pad{\'o}, S. 2026.
\newblock One {Persona}, {Many} {Cues}, {Different} {Results}: {How}
  {Sociodemographic} {Cues} {Impact} {LLM} {Personalization}.
\newblock \emph{arXiv preprint arXiv:2601.18572}.

\bibitem[{Whittle(2024)}]{TheConversation2024CopilotFalseAccusation}
Whittle, R. 2024.
\newblock Why Microsoft’s Copilot AI falsely accused court reporter of crimes
  he covered.
\newblock \emph{The Conversation}.
\newblock Accessed: 2026-05-22.

\bibitem[{Wilson et~al.(2025)Wilson, Sim, Gueorguieva, and
  Caliskan}]{wilson2025biased}
Wilson, K.; Sim, M.; Gueorguieva, A.-M.; and Caliskan, A. 2025.
\newblock No Thoughts Just AI: Biased LLM Hiring Recommendations Alter Human
  Decision Making and Limit Human Autonomy.
\newblock In \emph{Proceedings of the AAAI/ACM Conference on AI, Ethics, and
  Society}, volume~8, 2692--2704.

\bibitem[{Xu and Hu(2025)}]{xu2025rethinking}
Xu, S.~B.; and Hu, G. 2025.
\newblock Rethinking the author name ambiguity problem and beyond: The case of
  the Chinese context.
\newblock \emph{Accountability in research}, 32(6): 913--936.

\bibitem[{Xu et~al.(2025)Xu, Hu, Zhao, Qiu, Xu, Ye, and Gu}]{xu2025survey}
Xu, Y.; Hu, L.; Zhao, J.; Qiu, Z.; Xu, K.; Ye, Y.; and Gu, H. 2025.
\newblock A survey on multilingual large language models: Corpora, alignment,
  and bias.
\newblock \emph{Frontiers of Computer Science}, 19(11): 1911362.

\bibitem[{Ye et~al.(2024)Ye, Zhang, Zhou, Hu, Tian, and
  Cheng}]{ye2024correcting}
Ye, W.; Zhang, Q.; Zhou, X.; Hu, W.; Tian, C.; and Cheng, J. 2024.
\newblock Correcting Factual Errors in LLMs via Inference Paths Based on
  Knowledge Graph.
\newblock In \emph{2024 International Conference on Computational Linguistics
  and Natural Language Processing (CLNLP)}, 12--16. IEEE.

\bibitem[{Ye and Durrett(2022)}]{ye2022unreliability}
Ye, X.; and Durrett, G. 2022.
\newblock The unreliability of explanations in few-shot prompting for textual
  reasoning.
\newblock \emph{Advances in neural information processing systems}, 35:
  30378--30392.

\bibitem[{Zhang et~al.(2023)Zhang, Li, Hauer, Shi, and Kondrak}]{zhang2023don}
Zhang, X.; Li, S.; Hauer, B.; Shi, N.; and Kondrak, G. 2023.
\newblock Don’t Trust ChatGPT when your Question is not in English: A Study
  of Multilingual Abilities and Types of LLMs.
\newblock In \emph{Proceedings of the 2023 Conference on Empirical Methods in
  Natural Language Processing}, 7915--7927.

\bibitem[{Zhao et~al.(2025)Zhao, Zhang, Xu, and Wang}]{zhao2025surveyeval}
Zhao, J.; Zhang, S.; Xu, N.; and Wang, L. 2025.
\newblock SurveyEval: Towards Comprehensive Evaluation of LLM-Generated
  Academic Surveys.
\newblock \emph{arXiv preprint arXiv:2512.02763}.

\bibitem[{Zheng et~al.(2024)Zheng, Pei, Logeswaran, Lee, and
  Jurgens}]{zheng2024helpful}
Zheng, M.; Pei, J.; Logeswaran, L.; Lee, M.; and Jurgens, D. 2024.
\newblock When” a helpful assistant” is not really helpful: Personas in
  system prompts do not improve performances of large language models.
\newblock In \emph{Findings of the Association for Computational Linguistics:
  EMNLP 2024}, 15126--15154.

\end{thebibliography}
\end{document}